# Few-Shot Class-Incremental Audio Classification Using Pseudo-Incrementally Trained Embedding Learner and Continually Updated Stochastic Classifier

Yanxiong Li, *Senior Member*, *IEEE*, Wenchang Cao, Jiaxin Tan, Qianqian Li, and Guoqing Chen

***Abstract*—Few-shot Class-incremental Audio Classification (FCAC) aims to progressively recognize incremental classes with few tagged samples and meanwhile memorize base classes. To achieve satisfactory FCAC performance, the model needs to have high stability (memorizing base classes) and strong plasticity (adapting to incremental classes). In this work, we design a model which can be decoupled into two independent modules, namely an embedding learner and a stochastic classifier. The former is the backbone of a residual convolutional network, while the latter is composed of distributions and each distribution consists of a mean vector and a variance vector for representing one class. After being trained in the base session, the embedding learner is not updated in each incremental session and thus can memorize the knowledge of base classes. To make the embedding learner possess strong representation ability for incremental classes, we propose a strategy to pseudo-incrementally train the embedding learner using data augmentation in the base session. On the other hand, the stochastic classifier is continually updated in each incremental session and thus can adapt to incremental classes. Our model which consists of a pseudo-incrementally trained embedding learner and a continually updated stochastic classifier can increasingly identify incremental classes without forgetting base classes. Three datasets (FSC-89, NSynth-100 and LS-100) are used to verify the effectiveness of our method. Experiments show that our method exceeds the comparison methods in accuracy, and has lower complexity than most of the comparison methods. The code is at** https://github.com/vinceasvp/PITEL-CUSC.

## I. Introduction

AUDIO classification is to identify various kinds of sounds, which is a key step for many multimedia processing tasks, such as sound event detection [1]-[4], acoustic scene classification [5]-[8], video content analysis [9], [10], speaker diarization [11], [12] and speaker recognition [13], [14]. In practical situations, intelligent audio devices usually need to recognize classes that are not seen in the development stage (base session). To our best knowledge, the models that are transplanted on the intelligent audio devices can identify the predefined classes, but they are basically unable to continually learn unseen classes. The existing intelligent audio devices lack the ability to continually recognize new classes without forgetting old ones. To address this problem, some efforts are made in recent years.

### *A. Related Works*

The methods of audio classification can be divided into four categories, namely the methods based on supervised learning, few-shot learning, class-incremental learning, and few-shot class-incremental learning. They are briefly introduced as follows.

The method based on supervised learning is the most common and extensively studied method [15]-[21]. It requires a large number of training samples to recognize predefined classes, without the ability to identify new classes after the deployment of the trained model. In order for the model to recognize both new and old classes, it is required to retrain the model using abundant samples from all old and new classes. Retraining a model using abundant samples is time-consuming. In addition, owing to limited memory and data privacy, the samples of old classes are usually abandoned after the initial model training is completed. Hence, it is generally hard to obtain the samples of old classes again when retraining the model. If the model is updated (finetuned) by the samples of new classes only, it will catastrophically forget the knowledge of old classes and no longer have the ability to recognize old ones.

To reduce the amount of data for model training, the audio classification method based on few-shot learning is proposed [22]-[25]. Compared to the method based on supervised learning, the most prominent merit of the method based on few-shot learning is that very few training samples are required for model training. Moreover, the model can achieve satisfactory results after training with a small number of samples. The disadvantage of this method is that the model is inclined to overfitting to new classes and lacks the ability to memorize old classes. As new classes continue to emerge, knowledge of old ones will gradually be forgotten. Therefore, in this method, the model does not have the ability to memorize old classes.

To overcome the shortcomings of the above two types of methods that cannot memorize the knowledge of old classes, researchers propose the method based on class-incremental learning [26]-[28]. This method can identify new classes after training with abundant samples and memorize old classes in incremental sessions. Its disadvantage is that abundant samples of new classes are required to update the model for achieving satisfactory results. In many scenarios (collecting criminal's utterances and endangered animal's calls, etc.), it is not easy to acquire many samples of new classes owing to various reasons, such as data scarcity and high cost for data collection. Hence,

This work was partly supported by the national natural science foundation of China (62371195, 62111530145, 61771200), the exchange project of the 10th Meeting of the China-Croatia Science and Technology Cooperation Committee (No. 10-34) and Guangdong provincial key laboratory of human digital twin (2022B1212010004). *(Corresponding author: Yanxiong Li.)*

All authors of this paper are with the School of Electronic and Information Engineering, South China University of Technology, Guangzhou 510640, China. (e-mail: eeyxli@scut.edu.cn).

the method based on class-incremental learning is difficult to apply to scenarios where sample collection is challenging.

To overcome the deficiencies of the above three types of methods, some efforts are made on audio classification based on few-shot class-incremental learning [29]-[33]. In these works, the model is generally decoupled into two independent parts, i.e., an embedding learner and an expandable classifier. The embedding learner is trained in base session by abundant samples, and then is frozen in incremental sessions to avoid catastrophically forgetting the knowledge of base classes. On the contrary, the classifier is trained in base session with sufficient samples and updated in each incremental session to identify new classes with few training samples. Specifically, the audio classification method using dynamic few-shot learning (DFSL) is proposed for recognizing new classes without forgetting old ones [29]. A dynamically expandable classifier with self-attention modified prototypes is designed to implement the recognition of new classes and the memorization of old ones [30]. The expansible classifiers based on discriminative prototypes [31], adaptively refined prototypes [32], or fully-connected layers [33] are proposed for Few-shot Class-incremental Audio Classification (FCAC). In summary, the classifiers designed in these works generally consist of deterministic vectors (mean vectors) and belong to deterministic classifiers.

In the domains of computer vision and natural language processing, stochastic models (networks), especially Bayesian neural networks and their variants, are adopted for addressing some challenges like dataset shift, limited labeled data, and catastrophic forgetting. For example, the stochastic-YOLO [34] and deep stochastic adaptive Fourier decomposition network [35] are adopted for object detection and hyperspectral image classification, respectively. The above models have theoretical similarities with the stochastic classifier proposed in this paper. They adopt the probabilistic principles to estimate predictive uncertainty and improve generalization.

It should be noted that catastrophic forgetting is one of the critical issues addressed in incremental learning, but is not discussed in supervised learning and few-shot learning. In addition, some learning paradigms, such as transfer learning, semi-supervised learning and self-supervised learning, are adopted to solve the problem of audio classification. However, these learning paradigms cannot be directly used for solving the problem of FCAC.

### *B. Our Contributions*

Based on the above descriptions, it is known that the previous works can obtain satisfactory results of audio classification under different situations and undoubtedly promote research progress on the FCAC problem. In addition, previous methods are effective solutions for the two main issues confronted by the FCAC, namely overfitting to incremental classes and catastrophically forgetting of base classes. Each of previous methods has its unique strongpoints, but they still have the following two aspects that need to be further improved.

First, previous methods adopt abundant samples of base classes to train embedding learner in supervised learning way, without using samples of pseudo-incremental classes to train the embedding learner in few-shot learning way in base session. That is, previous methods aim to improve the representation ability of the embedding learner for base classes only, instead of for both base and incremental classes. As a result, the embedding learner trained by existing methods lacks the ability to effectively represent incremental classes.

Second, the classifier in prior methods generally consists of prototypes, and the weight of each class in the classifier is represented by a prototype. Each prototype of incremental class is the mean vector of embeddings of few support samples and is thus a deterministic vector. It is difficult for each deterministic vector to comprehensively characterize the incremental class, because the deterministic vector is learned from few support samples of one incremental class. In addition, support samples of each incremental class generally lack intra-class consistency, and each deterministic vector is less effective in representing one incremental class. Therefore, it is difficult for the deterministic classifier (consisting of deterministic vectors) to effectively distinguish all classes.

Inspired by the successes of data augmentation [36] and stochastic classifier [37] in the field of computer vision, we propose a FCAC method using a pseudo-incrementally trained embedding learner and a continually updated stochastic classifier. To train the embedding learner in few-shot learning way in base session, data augmentation is adopted to generate both the samples and labels of pseudo-incremental classes, instead of the samples only. The stochastic classifier is trained in base session and then updated in each incremental session to generate multiple weights for each class. Each weight is sampled from a distribution which is represented by multiple mean vectors and variance vectors. That is, a weight consists of a mean vector combined with a variance vector. Therefore, some representative weights around the mean vector can be generated and at least one of them is expected to effectively characterize each class. To extensively assess the performance of our method, we generate three datasets from different sources, namely the datasets of LS-100, FSC-89 and NSynth-100. These datasets are publicly available for research purpose, whose details will be introduced in the section of experiments (Section III). Experiments will verify the effectiveness of our method.

This work is an extension of our previous work published in Interspeech 2023 [38]. We add new content to our method and conduct more experiments. The differences between this work and our previous work in [38] mainly lie in three aspects. First, we propose a Pseudo-Incremental Training Strategy (PITS) to improve representation ability of the embedding learner for incremental classes. Second, we use three datasets to verify the effectiveness of our method, whereas only two datasets are adopted in [38]. Third, we do more experiments to evaluate the performance of our method and compare it with the state-of-the-art methods. In addition, we conduct a thorough analysis of the results.

In conclusion, the work in this paper has the following three contributions which are briefly introduced below.

1. We propose a strategy to pseudo-incrementally train the embedding learner using samples and labels generated by data augmentation in base session. After being trained by the augmented data, the embedding learner will have strong representation ability for both base and incremental classes, and thus the classification ability of the model will be improved. To the best of our knowledge, the proposed

PITS is novel and not considered in previous works.

2. We design a stochastic classifier which is dynamically updated in each incremental session for enhancing the classification performance of the model. The weights of the stochastic classifier are expressed in the form of distributions. Each weight is a mean vector combined with a variance vector. Hence, many weights around the mean vector can be generated and at least one of these weights is expected to represent each class well.
3. We propose a FCAC method using an embedding learner and a stochastic classifier. The effectiveness of our method is verified from several aspects on three public datasets. Extensive experiments show that our method outperforms previous methods in accuracy. In addition, it has advantage over most of previous methods in complexity.

In conclusion, the above three contributions make this work considerably different from previous works, and are beneficial for improving the FCAC performance. The rest of the paper is structured as follows. Section II presents our method in detail. Section III introduces extensive experiments and discussions. Finally, conclusions are made in Section IV.

## II. METHOD

In this section, we describe our method in detail, including the problem definition, method framework, embedding learner, stochastic classifier, PITS, and incremental training and inference of model.

### A. Problem Definition

As shown in Fig. 1, the task of FCAC consists of two kinds of sessions, base session (session 0) and incremental session (sessions 1 to $M$). $M$ denotes total number of incremental sessions. The training dataset $\boldsymbol{D}_0^{tr}$ of the base session is a large-scale dataset with $N_0$ classes and $K_0$ samples per class, while the training dataset $\boldsymbol{D}_m^{tr}$ ($1 \le m \le M$) of each incremental session is a small-scale dataset with $N_m$ classes and $K_m$ samples per class. Moreover, $N_0 >> N_m$ and $K_0 >> K_m$ ($1 \le m \le M$). The base and incremental sessions of FCAC task are equivalent to the development and application stages of intelligent audio devices, respectively. There are sufficient data for model training in the development stage, and few data for model update in the application stage.

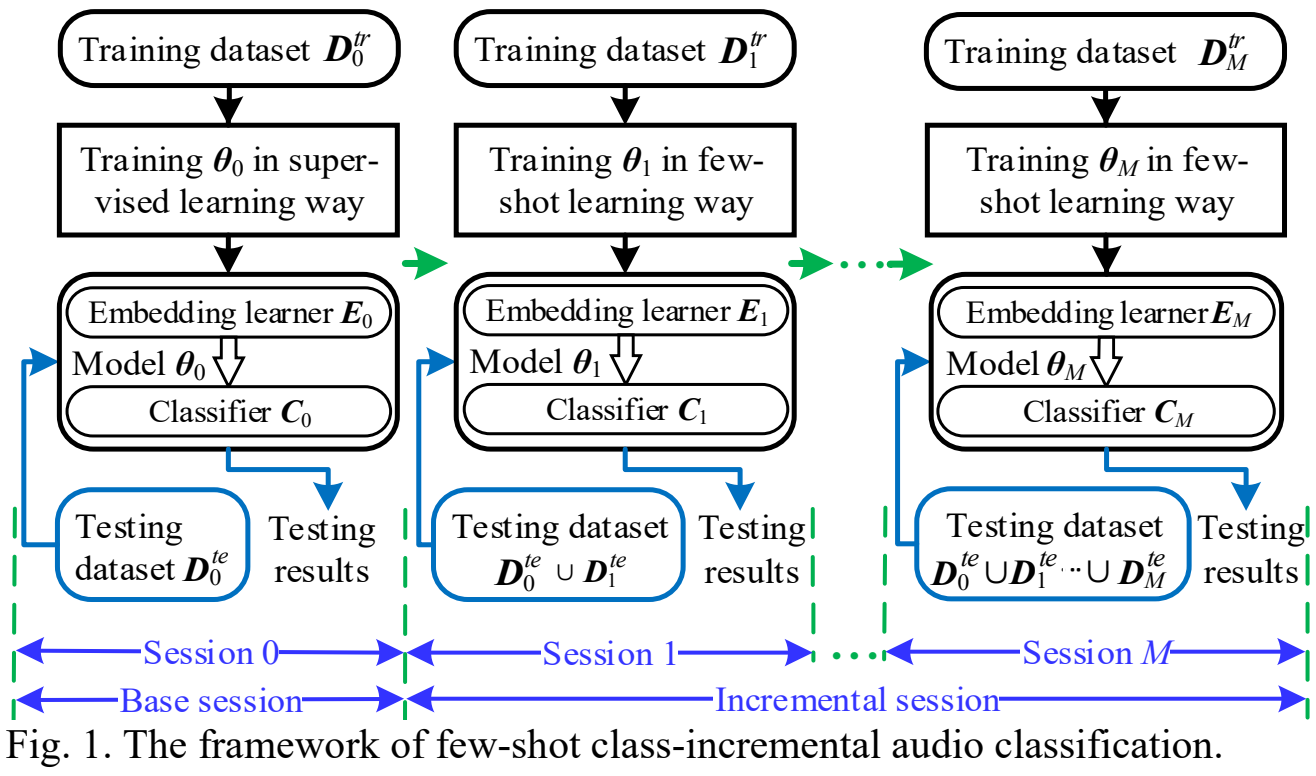

Fig. 1. The framework of few-shot class-incremental audio classification.

Each model consists of an embedding learner (e.g., $\boldsymbol{E}_0$) and a classifier (e.g., $\boldsymbol{C}_0$). The models of $\boldsymbol{\theta}_0$ and $\boldsymbol{\theta}_m$ ($1 \le m \le M$) are trained in supervised learning way with sufficient samples of base classes and in few-shot learning way with few samples of incremental classes, respectively. In the $m$th session, the training dataset $\boldsymbol{D}_m^{tr}$ only is available for training the model $\boldsymbol{\theta}_m$. However, In the $m$th session, the testing datasets of the $m$th session and all previous sessions, namely the testing datasets of $\boldsymbol{D}_0^{te} \cup \boldsymbol{D}_1^{te} \ldots \cup \boldsymbol{D}_m^{te}$ from all seen classes so far, need to be used for evaluating the performance of the model $\boldsymbol{\theta}_m$. The label sets of training dataset $\boldsymbol{D}_m^{tr}$ are the same as those of testing dataset $\boldsymbol{D}_m^{te}$ and are represented by $\boldsymbol{L}_m$. The datasets from different sessions do not have the same labels, namely $\forall l, m$ and $l \neq m, \boldsymbol{L}_l \cap \boldsymbol{L}_m = \phi$. The label sets of the $m$th session and all previous sessions are represented by $\boldsymbol{L}=\{\boldsymbol{L}_0 \cup \boldsymbol{L}_1 \ldots \cup \boldsymbol{L}_m\}$.

### B. Method Framework

As shown in Fig. 2, the framework for our method consists of one base session and $M$ incremental sessions. There are four consecutive steps in the base session, such as training model in supervised learning way, extracting embeddings, updating model by the PITS, and computing mean vectors. In the PITS, samples and labels of pseudo-incremental classes that are generated by the training data of base classes are used to train the model in a few-shot learning way (simulating the model training in the incremental sessions), and thus the model's ability to classify incremental classes is expected to be improved. Each incremental session has four steps, including extracting embeddings, computing mean vectors, combination, and updating stochastic classifier. In the base and incremental sessions, the audio feature of log Mel-spectrum [39] is extracted from each sample and fed into the embedding learner for extracting embedding. The log Mel-spectrum has been extensively adopted as the input of deep neural network for embedding extraction [40]-[42] and thus is also used here.

In the base session, a model $\boldsymbol{\theta}_0$ is first randomly initialized and then trained in a supervised learning way with abundant training samples of base classes. The trained model $\boldsymbol{\theta}_0$ consists of an embedding learner $\boldsymbol{E}_0$ and a stochastic classifier $\boldsymbol{C}_0$. Next, the trained model $\boldsymbol{\theta}_0$ is updated by the PITS in the base session for enhancing the generalization ability of the model for incremental classes, namely enhancing the model's plasticity. Finally, mean vector $\boldsymbol{\mu}_{0,n}$ of embeddings of the $n$th base class is computed. Mean vectors of all base classes are represented by $\boldsymbol{U}_0 = \{\boldsymbol{\mu}_{0,n}\}$, $1 \le n \le N_0$, where $N_0$ denotes total number of base classes. Mean vectors $\boldsymbol{U}_0$ are saved for further updating stochastic classifier in incremental session.

In the $m$th ($m \ge 1$) incremental session, the embedding learner which has been trained already in the base session is first adopted for extracting embeddings from log Mel-spectra of samples of all classes. Afterwards, mean vector $\boldsymbol{\mu}_{m,n}$ of embeddings of the $n$th class in the $m$th incremental session is computed and saved. Mean vectors of all classes of the $m$th incremental session are represented by $\boldsymbol{U}_m = \{\boldsymbol{\mu}_{m,n}\}$, $1 \le n \le N_m$, where $N_m$ represents total number of classes of the $m$th incremental session. Meanwhile, embeddings of all classes of the $m$th incremental session, and mean vectors of both base session and all $m$ incremental sessions are used to update the stochastic classifier. After finishing the updating procedure of the stochastic classifier, model $\boldsymbol{\theta}_m = \boldsymbol{E}_0 + \boldsymbol{C}_m$ is obtained.

It should be noted that embedding learner is frozen and stochastic classifier is continually updated in each incremental session. Our considerations for doing so are as follows. First, freezing the embedding learner is able to alleviate the model's forgetting of the knowledge of base classes. Second, we use the PITS to update the embedding learner, which enables the embedding learner to have strong generalization ability for incremental classes. Third, updating the stochastic classifier can improve the performance of the model for incremental classes.

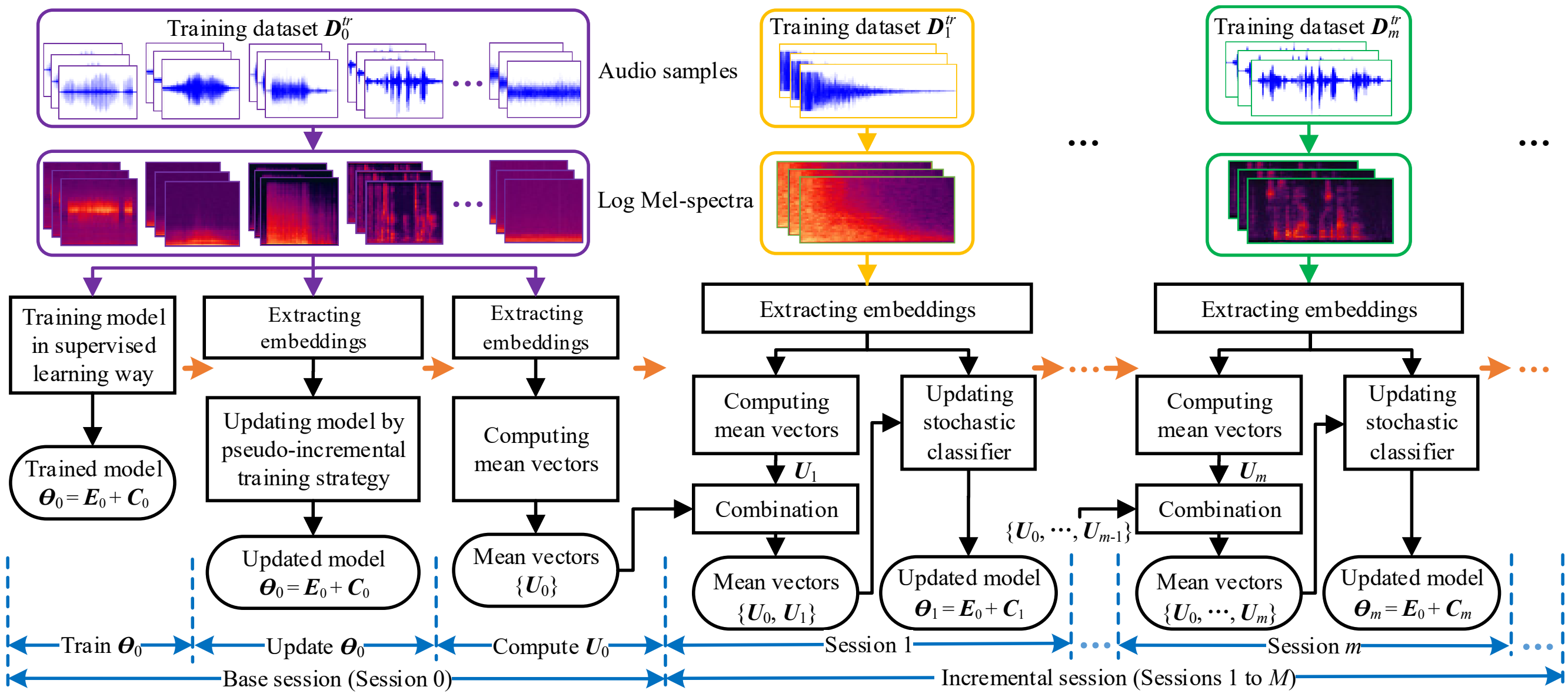

Fig. 2. Framework for our method. Model $\boldsymbol{\Theta}_0$ is first trained in supervised learning way with abundant samples and is then updated in few-shot learning way (pseudo-incremental training) in base session. Embedding learner $\boldsymbol{E}_0$ is frozen and stochastic classifier $\boldsymbol{C}_m$ is updated in few-shot learning way in the $m$th incremental session. $\boldsymbol{U}_0 = \{\boldsymbol{\mu}_{0,n}\}$, $1 \le n \le N_0$, and $\boldsymbol{\mu}_{0,n}$ denotes the mean vector of embeddings of the $n$th class in base session (session 0). $\boldsymbol{\mu}_{m,n}$ denotes the mean vector of embeddings of the $n$th class in the $m$th incremental session, and $\boldsymbol{U}_m = \{\boldsymbol{\mu}_{m,n}\}$, $1 \le n \le N_m$, $1 \le m \le M$.

In summary, the model is trained by abundant samples of base classes in supervised learning way, and updated by the PITS in the base session. Mean vector of each class in each session is computed and saved to update the stochastic classifier in next session. Mean vectors of embeddings are very compact, and can be seen as the result of severe lossy compression of audio samples. It is impossible to recover the audio samples from mean vectors. Hence, saving mean vectors for each class in each session does not violate the setting of FCAC problem and is without the risk of data privacy leakage. Moreover, its complexity is low. The embedding learner with plenty parameters is frozen in each incremental session to improve the model's stability for base classes, while the stochastic classifier with fewer parameters is updated by few samples of current incremental session to enhance the model's plasticity for incremental classes. Therefore, we not only solve the dilemma of model stability and plasticity, but also ensure that the computational complexity for model update is low.

### C. Embedding Learner

The backbone of deep convolutional neural network (DCNN) is generally used as the embedding learner, because it possesses strong ability to learn discriminative embeddings. The ResNet [43], a typical DCNN with residual connections, is popularly used for embeddings extraction in many tasks [42], [44], [45]. Inspired by the ResNet's success for extracting embeddings in many previous works, the embedding learner adopted in this work is the backbone of a ResNet18. The structure and parameters of the embedding learner used here is illustrated in Fig. 3.

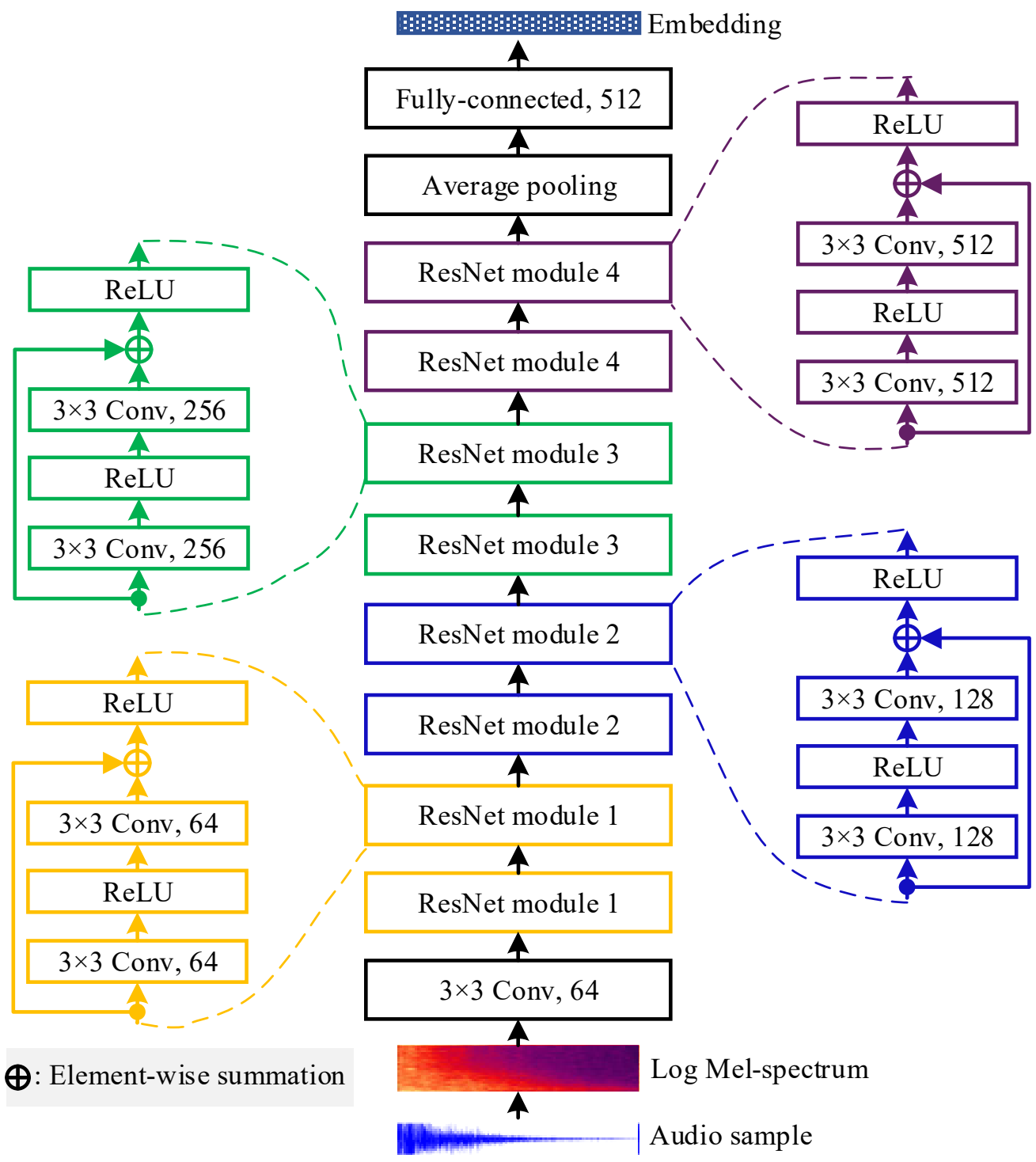

Fig. 3. Structure of the embedding learner.

We adopt a universal embedding learner with default settings of parameters, as our method can be applicable and effective for different embedding learners. The embedding learner is composed of one convolutional layer, eight ResNet modules with four different parameter settings, one layer of average pooling, and one fully-connected layer. It should be noted that a Softmax layer will be stacked on the fully-connected layer during the training procedure of the embedding learner. The Softmax layer will be discarded after finishing the training of the embedding learner. Each ResNet module is composed of two convolutional layers, two ReLU (Rectified Linear Unit) layers, and one skip connection. The parameters of each layer and module of the embedding learner are shown in Fig. 3. The embedding learner is trained under the supervision of cross-entropy loss. After completing the training procedure, the embedding is extracted from the fully-connected layer of embedding learner for each sample.

### *D. Stochastic Classifier*

The main considerations for choosing a stochastic classifier are twofold. First, the stochastic classifier is able to effectively distinguish incremental classes with only a small number of training samples, and thus overcome the shortcomings of the deterministic classifier (such as prototype classifier). Second, the stochastic classifier can alleviate the overfitting problem of the model to incremental classes.

A multivariate Gaussian distribution $\mathcal{N}(\boldsymbol{\mu}, \boldsymbol{\sigma})$ is constructed for generating the weights of a stochastic classifier, in which $\boldsymbol{\mu}$ and $\boldsymbol{\sigma}$ represent mean vector and variance vector, respectively. A vector $\hat{\boldsymbol{\mu}}$ is randomly sampled from the distribution $\mathcal{N}(\boldsymbol{\mu}, \boldsymbol{\sigma})$ and used as the weights of the stochastic classifier. The vector $\hat{\boldsymbol{\mu}}$ that is sampled from the distribution $\mathcal{N}(\boldsymbol{\mu}, \boldsymbol{\sigma})$ consists of stochastic variables, and thus cannot be directly updated by the backpropagation algorithm. Therefore, reparameterization technique [46] is introduced during the sampling process of the vector $\hat{\boldsymbol{\mu}}$ for updating parameters of the stochastic classifier, which decomposes the vector $\hat{\boldsymbol{\mu}}$ into a stochastic vector $\boldsymbol{\epsilon}$ and two trainable vectors of $\boldsymbol{\mu}$ and $\boldsymbol{\sigma}$, namely $\hat{\boldsymbol{\mu}}=\boldsymbol{\mu}+\boldsymbol{\epsilon}\odot\boldsymbol{\sigma}$. Vector $\boldsymbol{\epsilon}$ follows the distribution $\mathcal{N}(0, 1)$, namely $\boldsymbol{\epsilon}\sim\mathcal{N}(0, 1)$, and $\odot$ denotes the Hadamard product. The trainable vectors of $\boldsymbol{\mu}$ and $\boldsymbol{\sigma}$ are updated via the backpropagation algorithm. The reason why we choose a multivariate Gaussian distribution for generating the weights of the stochastic classifier is that it is the most commonly-used distribution in practical tasks with excellent performance and generality, convenient calculation, and complete ecosystem (e.g., rich algorithm library).

The $k$th sample of class $i$ and its embedding are represented by $\boldsymbol{x}_{i,k}$ and $f(\boldsymbol{x}_{i,k})$, respectively. The weight of a deterministic classifier for class $i$ is represented by $\boldsymbol{\mu}_i^d$ which is defined by

$$\boldsymbol{\mu}_i^d = \frac{1}{K}\sum_{k=1}^{K} f(\boldsymbol{x}_{i,k}) \ . \tag{1}$$

$\boldsymbol{\mu}_i^d$ is the mean vector of $K$ embeddings of the $i$th class. The cosine similarity between the embedding $f(\boldsymbol{x}_{i,k})$ and the weight $\boldsymbol{\mu}_i^d$ is defined by

$$\cos\left(f(\boldsymbol{x}_{i,k}),\ \boldsymbol{\mu}_i^d\right) = \frac{f(\boldsymbol{x}_{i,k})\boldsymbol{\mu}_i^d}{\|f(\boldsymbol{x}_{i,k})\|_2\|\boldsymbol{\mu}_i^d\|_2} , \tag{2}$$

where $\|\cdot\|_2$ stands for 2-norm. The Cosine similarity is widely used for the similarity measurement between embeddings (or feature vectors) in the field of audio processing. Hence, it is adopted here for measuring the similarity between embedding and weight. The weight of a stochastic classifier for class $i$ is denoted by $\hat{\boldsymbol{\mu}}_i = \boldsymbol{\mu}_i + \boldsymbol{\epsilon}_i \odot\boldsymbol{\sigma}_i$, where $\boldsymbol{\mu}_i$, $\boldsymbol{\epsilon}_i$ and $\boldsymbol{\sigma}_i$ represent mean vector, stochastic vector and variance vector for class $i$, respectively. The cosine similarity between the embedding $f(\boldsymbol{x}_{i,k})$ and the weight $\hat{\boldsymbol{\mu}}_i$ is defined by

$$\cos\left(f(\boldsymbol{x}_{i,k}),\ \hat{\boldsymbol{\mu}}_i\right) = \frac{f(\boldsymbol{x}_{i,k})\hat{\boldsymbol{\mu}}_i}{\|f(\boldsymbol{x}_{i,k})\|_2\|\hat{\boldsymbol{\mu}}_i\|_2} \ . \tag{3}$$

Fig. 4 depicts the difference between a stochastic classifier and a deterministic classifier. The area inside the dashed circle in Fig. 4 (a) or Fig. 4 (b) represents the embedding space of all classes. The center of the area where each class is located is the mean vector of the embeddings of this class, such as $\boldsymbol{\mu}_i^d$ (green arrow) and $\boldsymbol{\mu}_j^d$ (blue arrow) in Fig. 4 (a); $\boldsymbol{\mu}_i$ (green arrow) and $\boldsymbol{\mu}_j$ (blue arrow) in Fig. 4 (b). The classification boundary between any two classes is the bisector of the acute angle formed by the mean vectors of these two classes. For simplicity, we only show the mean vectors of two classes (class $i$ and class $j$) in both Fig. 4 (a) and Fig. 4 (b). Fig. 4 (a) illustrates two weights (namely mean vectors $\boldsymbol{\mu}_i^d$ and $\boldsymbol{\mu}_j^d$) and one embedding $f(\boldsymbol{x}_{i,k})$ for the deterministic classifier. The green diagonal area denotes the region where embedding $f(\boldsymbol{x}_{i,k})$ can be correctly identified as class $i$. $b_{ij}$ denotes the classification boundary between class $i$ (represented by the weight $\boldsymbol{\mu}_i^d$) and class $j$ (represented by the weight $\boldsymbol{\mu}_j^d$), which is a line (as shown in Fig. 4 (a)) since the deterministic classifier consists of mean vector only without variance vector. As illustrated in Fig. 4 (b), classification boundary between class $i$ (denoted by the weight $\hat{\boldsymbol{\mu}}_i$) and class $j$ (denoted by the weight $\hat{\boldsymbol{\mu}}_j$) turns into a margin (an area) which is caused by the variance vector (namely $\boldsymbol{\epsilon}_i \odot\boldsymbol{\sigma}_i$) of the stochastic classifier. The relationship between the margin and the variance vector is that the former is determined by the latter. The margins between any two classes are different as the variance vectors of the weights for any two classes are not the same.

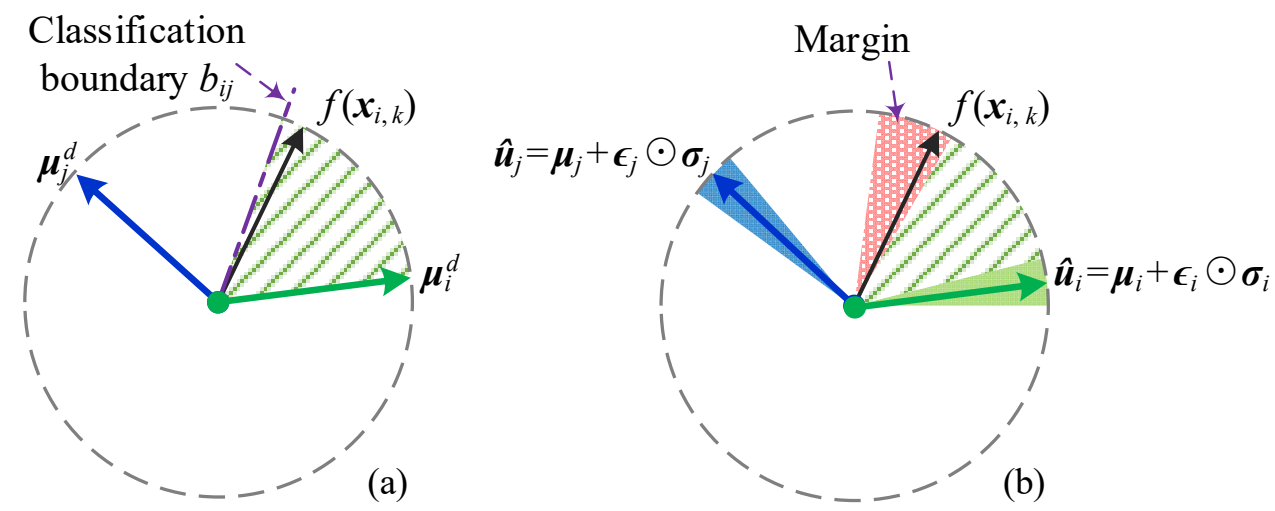

Fig. 4. Schematic diagrams of (a) a deterministic classifier and (b) a stochastic classifier. The classification boundaries between class $i$ and class $j$ in the deterministic and stochastic classifiers are a line (purple dashed line in (a)) and a margin (red grid area in (b)), respectively. The margin in (b) results in more discriminative classification boundaries between class $i$ and class $j$.

When $f(\boldsymbol{x}_{i,k})$ is within the margin, we can only have an ambiguous prediction that $f(\boldsymbol{x}_{i,k})$ probably belongs to class $j$. The above ambiguous prediction indicates that more powerful punishment should be put into effect in next training epoch and the embedding learner is enforced to generate more discriminative embeddings in base session. Furthermore, the stochastic classifier allows for distinguishing between the epistemic uncertainty and the aleatoric uncertainty [47]. Thanks to the above distinguishing feature of the stochastic classifier, incremental classes are expected to possess high epistemic uncertainty owing to the scarcity of their training samples. As a result, the overfitting problem of the model to incremental classes can be mitigated.

### *E. Pseudo-Incremental Training Strategy*

To prevent the model from forgetting the knowledge of base classes, the model trained on the samples of base classes should be frozen in incremental session. Freezing model in incremental session can enhance the stability of the model (memorizing of the knowledge of base classes). However, freezing model is not conducive to improving the plasticity of the model for incremental classes. To enhance the stability and plasticity of the model, we propose the PITS. Concretely, data augmentation is used to generate both samples and labels of pseudo-incremental classes based on the training data of base classes. The generated samples and labels are used to update the model in few-shot learning way. The updated embedding learner is expected to have strong representation ability for both base and incremental classes. In addition, the stochastic classifier is updated in each incremental session for enhancing the model's ability to distinguish incremental classes.

To make the generated samples conducive to the pseudo-incremental training of the model, we generate new samples and their labels of pseudo-incremental classes. Given two samples $\boldsymbol{x}_1$ and $\boldsymbol{x}_2$ and their labels $y_1$ and $y_2$, a new sample $\hat{\boldsymbol{x}}$ and its label $\hat{y}$ are obtained by Eq. (4) and Eq. (5), respectively.

$$\hat{\boldsymbol{x}} = \lambda\boldsymbol{x}_1 + (1\text{-}\lambda)\boldsymbol{x}_2, \tag{4}$$

$$\hat{y} = \lambda y_1 + (1\text{-}\lambda) y_2, \tag{5}$$

where mixing factor $\lambda$ is a coefficient ranging from 0 to 1. The label $\hat{y}$ is used to train the model just like the label of each base class. Namely, each pseudo incremental class is treated as an independent class like each base class. When λ is around 0 (or 1), $\hat{\boldsymbol{x}}$ is similar to the sample $\boldsymbol{x}_2$ (or $\boldsymbol{x}_1$). When λ is around 0.5, $\hat{\boldsymbol{x}}$ becomes a new sample that is different from both $\boldsymbol{x}_1$ and $\boldsymbol{x}_2$. $\lambda$ follows the Beta distribution with parameter of $\alpha$, namely $\lambda \sim$ Beta($\alpha$, $\alpha$). The probability density function of the Beta distribution is defined by

$$\varphi(\lambda;\alpha,\alpha) = \frac{\lambda^{\alpha-1}(1-\lambda)^{\alpha-1}}{\int_0^1 \lambda^{\alpha-1}(1-\lambda)^{\alpha-1} d\lambda}, \tag{6}$$

where $\alpha > 0$. The value of $\alpha$ determines the curve shape of the probability density function of the Beta distribution, and thus influences the probability of $\lambda$ taking different values.

To improve the model's ability for recognizing incremental classes, we design the PITS. Samples of both base and pseudo-incremental classes are used to train the model in the base session. Hence, the generalization ability of the model for incremental classes is expected to be enhanced. The label of pseudo-incremental class, $\hat{y}$, belongs to $\{N_0+1, \ldots, N_{\max}\}$, namely $\hat{y} \in \{N_0+1, \ldots, N_{\max}\}$, where $N_0$ denotes number of base classes, $N_{\max} = N_0 + M'N$ is total number of base classes and pseudo-incremental classes. $M'$ and $N$ represent the number of pseudo-incremental sessions and the number of classes in each pseudo-incremental session, respectively. Samples of pseudo-incremental classes are generated using samples of different base classes in the base session. In the process of pseudo-incremental training, training sample $\boldsymbol{x}$ can come from base classes or pseudo-incremental classes. Therefore, the model can learn knowledge of base classes and pseudo-incremental classes, and thus will not overfit to base classes.

From the above description, it can be known that data augmentation technique is used as a basic step in the PITS for generating samples and labels of pseudo-incremental classes. The PITS is a few-shot learning process. In each training episode, a few samples are randomly selected from training dataset of base classes to construct a support set ($N$ classes and $K$ samples per class) and a query set ($N$ classes and $L$ samples per class). That is, $N\times(K+L)$ samples are chosen in each training episode. The PITS is described in Table I. Its advantages and differentiation from prior augmentation-based incremental learning methods mainly lie in two aspects. First, it performs the operation of few-shot learning for simulating the few-shot learning in incremental sessions to update the model. Hence, it has low complexity and is easy to implement (only a few training samples are required for each class). Second, it uses samples and labels of both base and pseudo-incremental classes to update the model. As a result, the updated model will not overfit to base classes and will have strong generalization to incremental classes. In converse, the existing augmentation-based incremental learning methods do not perform the same operations in the above two aspects.

TABLE I
PSEUDO-INCREMENTAL TRAINING STRATEGY FOR UPDATING THE MODEL IN BASE SESSION.

**Input**:
Training dataset $\boldsymbol{D}_0^{tr}$, model $\boldsymbol{\theta}_0=\boldsymbol{E}_0+\boldsymbol{C}_0$ and $\boldsymbol{\theta}'_1=\boldsymbol{\theta}_0$. $\boldsymbol{\theta}_0$ and $\boldsymbol{\theta}'_1$ are models in base session and the first pseudo-incremental session, respectively.
**While** not done **do**:
Number of pseudo-incremental session $m$ is set to 1, namely $m = 1$.
**While** $m \le M'$:
1. Select a mixing factor $\lambda$ based on the beta distribution for generating samples and labels of pseudo-incremental classes.
2. Construct query set of the $m$th pseudo-incremental session, $\boldsymbol{Q}^{PI} = \{\boldsymbol{X}_{m,n}^{PI}; \boldsymbol{Y}_{m,n}^{PI}\}$ where $\boldsymbol{X}_{m,n}^{PI}=\{\hat{\boldsymbol{x}}_{m,n,l}^{PI}\}$ are query samples and $\boldsymbol{Y}_{m,n}^{PI}=\{\hat{y}_{m,n,l}^{PI}\}$ are labels of query samples, $1\le n\le N$, $1\le l\le L$. $\hat{\boldsymbol{x}}_{m,n,l}^{PI}=\lambda\boldsymbol{x}_1 + (1\text{-}\lambda)\boldsymbol{x}_2$ and $\hat{y}_{m,n,l}^{PI} = \lambda y_1 + (1\text{-}\lambda)y_2 \in \{N_0+(m\text{-}1)N+1, \ldots, N_0+mN\}$, where sample $\boldsymbol{x}_1$ with label $y_1$ is different from sample $\boldsymbol{x}_2$ with label $y_2$, and both of them are randomly selected from $\boldsymbol{D}_0^{tr}$.
3. Construct support set of the $m$th pseudo-incremental session, $\boldsymbol{S}^{PI} = \{\boldsymbol{X}_{m,n}^{PI}; \boldsymbol{Y}_{m,n}^{PI}\}$ where $\boldsymbol{X}_{m,n}^{PI}=\{\hat{\boldsymbol{x}}_{m,n,k}^{PI}\}$ are support samples and $\boldsymbol{Y}_{m,n}^{PI}=\{\hat{y}_{m,n,k}^{PI}\}$ are labels of support samples, $1\le n\le N$, $1\le k\le K$. $\hat{\boldsymbol{x}}_{m,n,k}^{PI}=\lambda\boldsymbol{x}_3 + (1\text{-}\lambda)\boldsymbol{x}_4$ and $\hat{y}_{m,n,k}^{PI} = \lambda y_3 + (1\text{-}\lambda)y_4 \in \{N_0+(m\text{-}1)N+1, \ldots, N_0+mN\}$, where sample $\boldsymbol{x}_3$ with label $y_3$ is different from sample $\boldsymbol{x}_4$ with label $y_4$, and both of them are randomly selected from $\boldsymbol{D}_0^{tr}$.
4. Compute mean vectors of pseudo-incremental classes, $\boldsymbol{U}^{PI} = \{\boldsymbol{\mu}_{m,n}^{PI}\}$, where $\boldsymbol{\mu}_{m,n}^{PI} = \frac{1}{K}\sum_{k=1}^{K} f(\hat{\boldsymbol{x}}_{m,n,k}^{PI})$ and $1\le n\le N$. $\hat{\boldsymbol{x}}_{m,n,k}^{PI}$ is the $k$th support sample of the $n$th pseudo-incremental class in the $m$th pseudo-incremental session, and $f(\hat{\boldsymbol{x}}_{m,n,k}^{PI})$ is embedding of $\hat{\boldsymbol{x}}_{m,n,k}^{PI}$.
5. The weight of stochastic classifier for the $n$th pseudo-incremental class in the $m$th pseudo-incremental session is initialized as mean vector $\boldsymbol{\mu}_{m,n}^{PI}$ as the weight of the stochastic classifier for each pseudo-incremental class mainly consists of a mean vector. The class number of stochastic classifier $\boldsymbol{C}'_m$ increases from $N_0+(m\text{-}1)N$ to $N_0+mN$.
6. Construct query set of base classes, $\boldsymbol{Q}^{B} = \{\boldsymbol{X}_{m,n}^{B}; \boldsymbol{Y}_{m,n}^{B}\}$, where $\boldsymbol{X}_{m,n}^{B}=\{\boldsymbol{x}_{m,n,l}^{B}\}$ are query samples and $\boldsymbol{Y}_{m,n}^{B}=\{y_{m,n,l}^{B}\}$ are labels of query samples, $1\le n\le N$, $1\le l\le L$.
7. Use the model $\boldsymbol{\theta}'_m = \boldsymbol{E}'_m + \boldsymbol{C}'_m$ of the $m$th pseudo-incremental session to predict the class label of each sample in both $\boldsymbol{Q}^{B}$ and $\boldsymbol{Q}^{PI}$, and calculate the cross-entropy loss $\mathcal{L}(\boldsymbol{Y}_{m,n}, \widetilde{\boldsymbol{Y}}_{m,n})$. $\boldsymbol{Y}_{m,n}$ and $\widetilde{\boldsymbol{Y}}_{m,n}$ denote the ground-truth and predicted labels, respectively.
8. Update parameters of the model $\boldsymbol{\theta}'_m$ by the algorithm of stochastic gradient descent based on the cross-entropy loss $\mathcal{L}(\boldsymbol{Y}_{m,n}, \widetilde{\boldsymbol{Y}}_{m,n})$.
9. $m = m + 1$.
**End while**
10. $\boldsymbol{\theta}'_1 = \boldsymbol{\theta}'_{M'}$.
**End while**
**Output**:
The updated $\boldsymbol{\theta}'_1 = \boldsymbol{E}'_1 + \boldsymbol{C}'_1$, and is assigned to $\boldsymbol{\theta}_0$, namely $\boldsymbol{\theta}_0 = \boldsymbol{\theta}'_1$.

### *F. Incremental Training and Inference of Model*

After being trained using the PITS in the base session, the embedding learner is expected to own strong representation ability for base and incremental classes. Hence, the embedding learner is frozen in incremental sessions. The stochastic classifier is updated for continually identifying incremental classes in each incremental session.

The model updating in the $m$th incremental session is a few-shot learning process. Weights of the stochastic classifier are initialized as the mean vectors of all previous $m$-1 sessions and the $m$th session. Then, the stochastic classifier is updated using mean vectors of all previous $m$-1 sessions and samples from $\boldsymbol{D}_m^{tr}$ which has $N$ classes with $K$ samples per class.

The loss for updating the stochastic classifier using training samples of the $m$th incremental session is defined by

$$\mathcal{L}(\boldsymbol{x}, y) = -\log(e^{\cos(f(\boldsymbol{x}),\ \widehat{\boldsymbol{\mu}}_y)} / \textstyle\sum_{h=1}^{|\boldsymbol{L}_m|} e^{\cos(f(\boldsymbol{x}),\ \widehat{\boldsymbol{\mu}}_h)}) \ , \quad (7)$$

where $\boldsymbol{x}$ and $y$ are a training sample and its label, respectively. $\boldsymbol{L}_m$ denotes the labels in the $m$th incremental session, and $|\boldsymbol{L}_m|$ is the number of elements in $\boldsymbol{L}_m$. $\widehat{\boldsymbol{\mu}}_y = \boldsymbol{\mu}_y + \mathcal{N}(0,1) \odot \boldsymbol{\sigma}_y$ represents the weight of stochastic classifier for the $y$th class. $f(\boldsymbol{x})$ denotes the embedding of $\boldsymbol{x}$. $\boldsymbol{\mu}_y$ and $\boldsymbol{\sigma}_y$ represent the mean vector and variance vector of the embeddings of the $y$th class, respectively. $\cos(f(\boldsymbol{x}),\ \widehat{\boldsymbol{\mu}}_y)$ stands for the cosine similarity between the embedding of training sample $\boldsymbol{x}$ and the weight of stochastic classifier for the $y$th class.

The loss for updating the stochastic classifier using mean vectors of all previous $m$-1 sessions is defined by

$$\mathcal{L}(\boldsymbol{\mu}, y') = -\log(e^{\cos(\boldsymbol{\mu},\ \widehat{\boldsymbol{\mu}}_y)} / \textstyle\sum_{h=1}^{|\boldsymbol{L}'|} e^{\cos(\boldsymbol{\mu},\ \widehat{\boldsymbol{\mu}}_h)}) \ , \quad (8)$$

where $\boldsymbol{\mu}$ and $y'$ denote a mean vector of previous $m$-1 sessions and its label, respectively; $\boldsymbol{L}' = \boldsymbol{L}_0 \cup \boldsymbol{L}_1 \cdots \cup \boldsymbol{L}_{m-1}$; and $|\boldsymbol{L}'|$ stands for the number of elements in $\boldsymbol{L}'$. The total loss for updating the stochastic classifier is defined by

$$\mathcal{L} = \mathcal{L}(\boldsymbol{x}, y) + \mathcal{L}(\boldsymbol{\mu}, y') \ . \quad (9)$$

During inference in the $m$th session, the predicted class label $\tilde{y}$ for a testing sample $\boldsymbol{x}$ is obtained by

$$\tilde{y} = arg \max_{y} \frac{e^{cos(f(x),\ \widehat{\mu}_y)}}{\sum_{h=1}^{|L|} e^{cos(f(x),\ \widehat{\mu}_h)}} \ . \quad (10)$$

where $\boldsymbol{L} = \boldsymbol{L}_0 \cup \cdots \cup \boldsymbol{L}_m$, and $|\boldsymbol{L}|$ is the number of elements in $\boldsymbol{L}$.

## III. EXPERIMENTS

In this section, we first introduce experimental datasets and setup. Then, we present results and discussions of ablation, comparative and extended experiments.

### A. Experimental Datasets

The FSC-89, NSynth-100 and LS-100 are used as experimental datasets to assess the performance of different methods, which are widely used in previous works for audio classification. To facilitate reimplementation of our results, the details of these three datasets are publicly available at three websites[1].

The FSC-89 has 89 sound events which cover a diverse range of real-world sounds, from human and animal sounds to natural, musical or miscellaneous sounds. The NSynth-100 has 100 sounds of musical instruments. The LS-100 consists of speech samples uttered by 100 different speakers. The details of the original audio corpora for the FSC-89, NSynth-100 and LS-100 can be obtained in [48], [49] and [50], respectively.

Each dataset is divided into two subsets: base subset $\boldsymbol{D}_0$ and incremental subset $\boldsymbol{D}_m$ ($1 \le m \le M$). These two subsets are further divided into training dataset $\boldsymbol{D}_0^{tr}$ (or $\boldsymbol{D}_m^{tr}$) and testing dataset $\boldsymbol{D}_0^{te}$ (or $\boldsymbol{D}_m^{te}$). The $\boldsymbol{D}_0^{tr}$ is used to train the model (consisting of an embedding learner and a stochastic classifier) in supervised learning way, and is used to generate samples of pseudo-incremental classes. Then, the samples randomly selected from both $\boldsymbol{D}_0^{tr}$ and generated pseudo-incremental dataset are used to update the model in each pseudo-incremental session.

[1] https://www.modelscope.cn/datasets/pp199124903/LS-100/summary
https://www.modelscope.cn/datasets/pp199124903/FSC-89/summary
https://www.modelscope.cn/datasets/pp199124903/NSynth-100/summary

In the $m$th incremental session, the $\boldsymbol{D}_m^{tr}$ ($1 \le m \le M$) is used to update the stochastic classifier in few-shot learning way. In each training episode, $N \cdot K$ and $N \cdot L$ samples are randomly chosen from the $\boldsymbol{D}_m^{tr}$ as the support and query sets, respectively. Testing samples of all previous sessions and the $m$th (current) session are used to assess the model's performance. The details for data divisions of the three datasets are listed in Table II.

TABLE II
DATA DIVISIONS OF THE THREE DATASETS

| Datasets | Parameters | $\boldsymbol{D}_0$ | | $\boldsymbol{D}_m$ | |
|---|---|---|---|---|---|
| | | $\boldsymbol{D}_0^{tr}$ | $\boldsymbol{D}_0^{te}$ | $\boldsymbol{D}_m^{tr}$ | $\boldsymbol{D}_m^{te}$ |
| FSC-89 | #Classes | 59 | 59 | 30 | 30 |
| | #Samples | 47200 | 11800 | 15000 | 6000 |
| | Length (hours) | 13.11 | 3.28 | 4.17 | 1.67 |
| | #Samples/Class | 800 | 200 | 500 | 200 |
| NSynth-100 | #Classes | 55 | 55 | 45 | 45 |
| | #Samples | 11000 | 5500 | 4500 | 4500 |
| | Length (hours) | 12.23 | 1.52 | 5.00 | 5.00 |
| | #Samples/Class | 200 | 100 | 100 | 100 |
| LS-100 | #Classes | 60 | 60 | 40 | 40 |
| | #Samples | 30000 | 6000 | 20000 | 4000 |
| | Length (hours) | 16.66 | 3.33 | 11.11 | 2.22 |
| | #Samples/Class | 500 | 100 | 500 | 100 |

#Samples/Class: number of samples for each class; $\boldsymbol{D}_m$: audio samples of all classes in all incremental sessions, $1 \le m \le M$.

### B. Experimental Setup

Experiments are done on a computer with two Intel CPUs of Xeon-8124M with 3.50 GHz, a RAM with 128 GB, and three NVIDIA 3090 GPUs. Two dominant metrics of Average accuracy (AA) and performance dropping rate (PD) are used to measure the performance of all methods. AA is defined by

$$AA = \frac{1}{M+1} \textstyle\sum_{m=0}^{M} A_m, \quad (11)$$

where $A_m$ is the accuracy in the $m$th session. PD is defined by

$$PD = A_0 - A_M \ . \quad (12)$$

Compared to PD, AA is an average amount and thus can more effectively represent performance of various methods. AA is adopted as primary metric. Multiply-ACcumulate operations (MACs), Training Time (TT) and Inference Time (IT) for one sample are adopted to measure the computational complexity. MACs represents the number of multiplications and additions required by various methods to complete one inference on one sample. TT and IT stand for the specific time required for various methods to complete one training and one inference on one sample. When experimental platform (such as computer hardware) changes, the value of MACs will not change, but the values of both TT and IT will change accordingly. Number of Parameters (NP) during inference is used to measure the memory overhead, which can be also used to represent the number of trainable parameters of various methods. Main parameters of our method are listed in Table III.

TABLE III
CONFIGURATIONS OF MAIN PARAMETERS OF OUR METHOD

| Parameters | Settings |
|---|---|
| Frame length/overlapping: | 25ms/10ms |
| Dimension of log Mel-spectrum: | 128 |
| Dimension of embedding, $D$: | 512 |
| Learning rate: | 0.1 (embedding learner), 0.0002 (classifier) |
| Batch size: | 128 |
| Weight decay: | 0.0005 |
| $\alpha$ values of the Beta distribution: | 0.4 (FSC-89), 3.0 (NSynth-100), 4.0 (LS-100) |
| No. of ways ($N$), shots ($K$ and $L$): | 1 to 20, 1 to 20 |
| No. of pseudo-incremental session: | 1 to 10 |

TABLE IV
RESULTS OBTAINED BY OUR METHOD ON THE LS-100 WITH DIFFERENT COMBINATIONS OF PSEUDO-INCREMENTAL TRAINING STRATEGY AND CLASSIFIERS

| Case | PITS | DC/SC | Accuracy scores in different sessions (%) | | | | | | | | | AA (%) | PD (%) |
|---|---|---|---|---|---|---|---|---|---|---|---|---|---|
| | | | 0 (0-59) | 1 (60-64) | 2 (65-69) | 3 (70-74) | 4 (75-79) | 5 (80-84) | 6 (85-89) | 7 (90-94) | 8 (95-99) | | |
| ① | × | DC | 91.85±0.73 | 91.84±0.82 | 91.20±0.52 | 88.65±0.45 | 86.81±0.62 | 86.02±0.60 | 84.88±0.82 | 82.31±0.61 | 81.75±0.80 | 87.26±0.66 | 10.10±0.07 |
| ② | × | SC | 92.06±0.64 | 91.69±0.54 | 91.44±0.67 | 89.41±0.53 | 88.12±0.55 | 87.17±0.58 | 86.29±0.46 | 83.72±0.75 | 82.77±0.69 | 88.08±0.60 | 9.29±0.05 |
| ③ | √ | DC | 92.05±0.36 | 91.94±0.50 | 91.29±0.71 | 89.53±0.36 | 88.61±0.69 | 87.89±0.45 | 87.07±0.52 | 84.74±0.49 | 84.29±0.58 | 88.60±0.52 | 7.76±0.22 |
| ④ | √ | SC | 92.03±0.81 | 92.41±0.64 | 92.31±0.65 | 90.77±0.27 | 90.36±0.58 | 89.60±0.42 | 88.70±0.67 | 86.70±0.64 | 86.17±0.49 | **89.90**±0.57 | **5.86**±0.32 |

×/√: the module isn't/is adopted. 0 (0-59): session 0 whose class numbers are from 0 to 59.

By setting a proper $\alpha$ value based on the training data from each experimental dataset, our method can achieve better results. Based on different training data from the three experimental datasets, $\alpha$ is set to 0.4, 3.0 and 4.0 for the FSC-89, NSynth-100 and LS-100, respectively. Main parameters of prior methods are set according to the suggestions of their original papers and tuned on the three experimental datasets.

### C. Ablation Experiments

In this subsection, we conduct ablation experiments to show effectiveness of each main module of our method. For simplicity, only the LS-100 is adopted and $N$-way $K$-shot is set to 5-way 5-shot for few-shot learning in ablation experiments.

As illustrated in Fig. 2, novelties of our method are two-fold. Namely, we propose the PITS and design a continually-updated stochastic classifier. To show the contributions of the above two modules to the performance improvement obtained by our method, we implement the FCAC method with the stochastic classifier (SC) or with deterministic classifier (DC), and with or without the PITS. Results obtained by our method on the LS-100 with different combinations of the PITS and the DC/SC are listed in Table IV.

From the results of case ① and case ②, we can know that our method with SC obtains AA score of (88.08±0.60)% and PD score of (9.29±0.05)% which are higher and lower than the AA score of (87.26±0.66)% and PD score of (10.10±0.07)%, respectively. From the results of case ③ and case ④, we can see that our method with SC obtains AA score of (89.90±0.57)% and PD score of (5.86±0.32)% which are also higher and lower than the AA score of (88.60±0.52)% and PD score of (7.76±0.22)%, respectively. In short, regardless of whether the PITS is used or not, our method achieves better performance when using the SC instead of the DC.

Based on the results of case ① and case ③, it can be seen that our method with the PITS obtains AA score of (88.60±0.52)% and PD score of (7.76±0.22)% which are higher and lower than the AA score of (87.26±0.66)% and PD score of (10.10±0.07)%, respectively. Based on the results of case ② and case ④, it can be observed that our method with the PITS obtains AA score of (89.90±0.57)% and PD score of (5.86±0.32)% which are also higher and lower than the AA score of (88.08±0.60)% and PD score of (9.29±0.05)%, respectively. In conclusion, whether the DC or SC is used, the performance of our method is improved by the PITS.

From the results in Table IV, we can conclude that the performance of our method is improved by using the PITS or SC separately. When the PITS and SC are used, our method achieves the best performance. Therefore, the PITS and SC are complementary for improving the performance of our method.

### D. Comparison in Accuracy

In this subsection, we implement comparison experiments to compare our method with twelve previous FCAC methods in accuracy. Their abbreviations are as follows: Finetune-C [51], Finetune-M [51], iCaRL [52], DFSL [29], CEC [53], FACT [54], CLOM [55], LDC [56], DSN [57], ARP [32], PAN [30], and AMFO [33]. The finetuning technique is a typical practice of model updating, which can enable the model to classify new classes. In this paper, we take the method of [51] as a representative of finetuning methods.

In the Finetune-C method, the backbone (i.e., embedding extractor) of the model is frozen and only the classifier of the model is updated in each incremental session. As a result, the incremental classes of current session can be classified well due to the update of the classifier, and the knowledge of the base classes will not be completely forgotten because of the freezing of the backbone. In the Finetune-M method, the model (consisting of the backbone and classifier) is updated with samples of incremental classes in each incremental session, and thus incremental classes of current session can be classified well. Nevertheless, the knowledge of both the base classes and all previous incremental classes is almost forgotten due to continual finetuning of all parameters of the model and thus these classes are difficult to be classified well. In the iCaRL method, both the knowledge distillation and data retention are adopted to update the model for obtaining satisfactory classification results. In the DFSL method, a weight generator with attention block is designed to generate weights of the classifier for identifying different classes. The classifier makes a decision based on the cosine similarity between its weights and the embedding of a testing sample.

In the CEC method, the model consists of two decoupled parts, namely embedding learner and expandable classifier. The former is trained in base session and then frozen in incremental sessions, whereas the latter is continually updated by a graph neural network in incremental sessions. In the FACT method, the embedding space of incremental classes is reserved in base session with the aim of reducing the confusions among different classes. In the CLOM method, one Softmax loss with a tunable positive margin and one Softmax loss with a tunable negative margin are independently used to train two classifiers. The final prediction result is the linearly weighted sum of the outputs of these two classifiers.

In the LDC method, a strategy of learnable distribution calibration is proposed to initialize and gauge the distributions of all classes, in which model parameters can be reused for identifying incremental classes. In the DSN method, the embedding space is constructed by a dynamic support network with dilatation of neural nodes, and the distributions of old classes can be recalled in current session. In the ARP and PAN methods, a model consists of an embedding learner and a deterministic classifier. The classifier consists of prototypes (mean vectors of embeddings of the same class) and each class is denoted by a prototype. A module of dynamic relation projection in the ARP method and a self-attention module in the PAN method are used to update all prototypes so far in each incremental session. In the AMFO method, the model consists of an embedding learner and an expandable classifier. The former is the backbone of a residual neural network, while the latter consists of two sub-classifiers and one fusion module.

Table V presents technical features of different methods. We implement previous methods using the code provided by the original papers or the code written by us according to the introductions in their respective papers. We configure main parameters of the previous methods based on the descriptions in the original papers and optimally adjusted on experimental data. The performance of all methods is assessed on three datasets. $N$-way $K$-shot is set to 5-way 5-shot for few-shot learning in experiments of this subsection. Experimental results obtained by different methods on the FSC-89, NSynth-100 and LS-100 are presented in Tables VI, VII, and VIII, respectively.

TABLE V
TECHNICAL FEATURES OF DIFFERENT METHODS

| Methods | Main features |
|---|---|
| Finetune-C | Update parameters of classifier with incremental-classes' samples |
| Finetune-M | Update parameters of model with incremental-classes' samples |
| iCaRL | Knowledge distillation and data retention |
| DFSL | Attentive weight generator; cosine-similarity based classifier |
| CEC | Decoupled model (an embedding learner and a classifier) |
| FACT | Reserve embedding space of incremental classes in base session |
| CLOM | Fusion of two classifiers with positive and negative margins |
| LDC | Gauge all-classes' distributions by learnable distribution calibration |
| DSN | A network with node dilatation; distribution recall of old classes |
| ARP | Update prototypes by a module of dynamic relation projection |
| PAN | A self-attention module is designed to update all prototypes so far |
| AMFO | Generalizable and discriminative embedding learners & classifiers |
| Ours | Pseudo-incremental training strategy and stochastic classifier |

TABLE VI
RESULTS OBTAINED BY DIFFERENT METHODS ON THE FSC-89

| Methods | Accuracy scores in different sessions (%) | | | | | | | AA (%) | PD (%) |
|---|---|---|---|---|---|---|---|---|---|
| | 0 (0-58) | 1 (59-63) | 2 (64-68) | 3 (69-73) | 4 (74-78) | 5 (79-83) | 6 (84-88) | | |
| Finetune-C [51] | 42.28 | 32.12 | 30.01 | 25.72 | 22.20 | 21.02 | 17.89 | 27.32 | 24.39 |
| Finetune-M [51] | 42.28 | 31.67 | 29.11 | 24.63 | 21.74 | 20.17 | 16.70 | 26.61 | 25.58 |
| iCaRL [52] | 42.48 | 32.40 | 30.25 | 25.99 | 23.42 | 19.78 | 19.44 | 27.68 | 23.04 |
| DFSL [29] | 42.36 | 35.23 | 32.72 | 31.21 | 29.79 | 28.51 | 27.40 | 32.46 | 14.96 |
| CEC [53] | 42.16 | 39.84 | 36.88 | 35.63 | 34.94 | 34.15 | 32.96 | 36.65 | 9.20 |
| FACT [54] | 43.14 | 39.76 | 36.41 | 34.85 | 33.30 | 31.32 | 30.66 | 35.63 | 12.48 |
| CLOM [55] | 37.61 | 35.06 | 32.99 | 31.44 | 30.23 | 29.13 | 28.04 | 32.07 | 9.57 |
| LDC [56] | 38.85 | 36.31 | 34.41 | 32.80 | 31.54 | 30.48 | 28.91 | 33.33 | 9.94 |
| DSN [57] | 38.99 | 36.57 | 34.01 | 32.49 | 30.68 | 29.16 | 27.91 | 32.83 | 11.08 |
| ARP [32] | 42.04 | 39.95 | 37.01 | 34.68 | 32.97 | 31.45 | 30.09 | 35.46 | 11.95 |
| PAN [30] | 42.92 | 40.01 | 37.84 | 37.27 | 36.73 | 35.88 | 34.71 | 37.91 | 8.21 |
| AMFO [33] | 43.97 | 41.22 | 39.46 | 38.42 | 38.04 | 37.16 | 36.04 | 39.19 | **7.93** |
| Ours | **46.89** | **44.78** | **42.06** | **41.07** | **40.53** | **39.44** | **38.40** | **41.88** | 8.49 |

TABLE VII
RESULTS OBTAINED BY DIFFERENT METHODS ON THE NSYNTH-100

| Methods | Accuracy scores in different sessions (%) | | | | | | | | | | AA (%) | PD (%) |
|---|---|---|---|---|---|---|---|---|---|---|---|---|
| | 0 (0-54) | 1 (55-59) | 2 (60-64) | 3 (65-69) | 4 (70-74) | 5 (75-79) | 6 (80-84) | 7 (85-89) | 8 (90-94) | 9 (95-99) | | |
| Finetune-C [51] | 99.96 | 86.12 | 79.43 | 74.61 | 66.23 | 43.21 | 50.31 | 44.21 | 42.23 | 40.37 | 62.67 | 59.59 |
| Finetune-M [51] | 99.96 | 84.73 | 76.92 | 71.06 | 63.18 | 40.15 | 47.99 | 38.33 | 38.01 | 37.86 | 59.82 | 62.10 |
| iCaRL [52] | 99.98 | 93.30 | 88.88 | 84.82 | 79.57 | 67.65 | 65.92 | 59.65 | 54.62 | 51.01 | 74.54 | 48.97 |
| DFSL [29] | 99.93 | 96.00 | 92.95 | 89.26 | 86.47 | 83.66 | 80.28 | 77.68 | 76.12 | 75.01 | 85.74 | 24.92 |
| CEC [53] | 99.96 | 97.47 | 95.56 | 93.52 | 91.22 | 89.90 | 87.85 | 85.84 | 84.92 | 83.19 | 90.94 | 16.77 |
| FACT [54] | 99.73 | 96.52 | 94.52 | 92.96 | 90.37 | 89.37 | 86.48 | 85.01 | 84.41 | 82.98 | 90.24 | 16.75 |
| CLOM [55] | 99.91 | 96.55 | 94.42 | 92.26 | 90.51 | 89.80 | 87.75 | 86.41 | 85.40 | 83.74 | 90.67 | 16.17 |
| LDC [56] | 99.71 | 97.32 | 94.09 | 92.57 | 90.16 | 89.49 | 87.85 | 86.71 | 86.25 | 84.87 | 90.90 | 14.84 |
| DSN [57] | **100** | 98.05 | 97.32 | 95.27 | 93.56 | 92.78 | 91.16 | 90.53 | 90.05 | 88.89 | 93.76 | 11.11 |
| ARP [32] | 99.96 | 95.95 | 93.60 | 92.06 | 90.32 | 89.14 | 86.16 | 83.47 | 82.28 | 79.69 | 89.26 | 20.27 |
| PAN [30] | 99.98 | 98.32 | 97.69 | 95.18 | 94.00 | 92.73 | 90.59 | 89.24 | 88.28 | 87.08 | 93.31 | 12.90 |
| AMFO [33] | 99.95 | 98.92 | 97.54 | 96.54 | 95.25 | 94.31 | 92.52 | 92.26 | 92.03 | 91.00 | 95.03 | 8.95 |
| Ours | 99.98 | **99.00** | **97.77** | **96.58** | **95.43** | **94.93** | **93.39** | **92.42** | **92.13** | **91.77** | **95.34** | **8.21** |

TABLE VIII
RESULTS OBTAINED BY DIFFERENT METHODS ON THE LS-100

| Methods | Accuracy scores in different sessions (%) | | | | | | | | | AA (%) | PD (%) |
|---|---|---|---|---|---|---|---|---|---|---|---|
| | 0 (0-59) | 1 (60-64) | 2 (65-69) | 3 (70-74) | 4 (75-79) | 5 (80-84) | 6 (85-89) | 7 (90-94) | 8 (95-99) | | |
| Finetune-C [51] | 92.02 | 76.23 | 39.32 | 31.18 | 24.78 | 19.61 | 12.94 | 9.58 | 7.88 | 34.84 | 84.14 |
| Finetune-M [51] | 92.02 | 73.95 | 36.24 | 28.27 | 21.93 | 17.33 | 9.86 | 7.00 | 4.88 | 32.39 | 87.14 |
| iCaRL [52] | 92.02 | 79.05 | 72.31 | 58.24 | 25.23 | 16.80 | 31.83 | 28.54 | 27.41 | 47.94 | 64.61 |
| DFSL [29] | 91.93 | 88.97 | 87.60 | 83.61 | 81.11 | 80.01 | 77.22 | 74.26 | 73.25 | 81.99 | 18.68 |
| CEC [53] | 91.72 | 91.25 | 90.04 | 86.84 | 85.38 | 84.01 | 82.65 | 79.49 | 78.45 | 85.54 | 13.27 |
| FACT [54] | 90.85 | 87.45 | 83.94 | 80.69 | 77.85 | 76.11 | 74.53 | 71.75 | 70.07 | 79.25 | 20.78 |
| CLOM [55] | 91.80 | 90.52 | 87.46 | 84.45 | 81.89 | 80.80 | 79.39 | 76.65 | 75.40 | 83.15 | 16.40 |
| LDC [56] | 92.28 | 90.95 | 88.53 | 85.64 | 83.67 | 82.25 | 81.11 | 78.75 | 77.64 | 84.54 | 14.64 |
| DSN [57] | 92.43 | 92.12 | 89.37 | 86.39 | 84.02 | 83.31 | 81.82 | 78.85 | 78.28 | 85.18 | 14.15 |
| ARP [32] | 92.35 | 88.43 | 85.16 | 81.53 | 78.22 | 75.73 | 72.84 | 69.87 | 67.54 | 79.07 | 24.81 |
| PAN [30] | 91.83 | 91.29 | 90.70 | 88.73 | 86.42 | 85.09 | 83.41 | 80.84 | 79.24 | 86.39 | 12.59 |
| AMFO [33] | **92.63** | 92.25 | 92.20 | 90.17 | 88.75 | 88.32 | 87.90 | 85.97 | 84.99 | 89.24 | 7.64 |
| Ours | 92.03 | **92.41** | **92.31** | **90.77** | **90.36** | **89.60** | **88.70** | **86.70** | **86.17** | **89.90** | **5.86** |

From the results in Tables VI to VIII, it can be seen that our method obtains AA scores of 41.88%, 95.34%, and 89.90% on the FSC-89, NSynth-100, and LS-100, respectively. These three AA scores obtained by our method are higher than the counterparts achieved by all previous methods. Moreover, our method obtains PD scores of 8.49%, 8.21%, and 5.86% on the FSC-89, NSynth-100, and LS-100, respectively. It outperforms most of previous methods in PD scores. In short, our method exceeds previous methods in accuracy on the three datasets. In addition, the AA scores obtained by all methods on the FSC-89 are consistently lower than those on the other two datasets, since the samples in the FSC-89 contain evident noises and the intra-class consistency of the samples in the FSC-89 is poor.

The above accuracy gains achieved by our method is mainly ascribed to the usages of the PITS and stochastic classifier. The above strategy can improve the representation ability of the embedding learner for all classes. The embedding learner is frozen in the incremental session, which enables the model to have high stability for base classes and strong plasticity for incremental classes. The stochastic classifier can effectively represent the character differences of various classes, and is continually updated in each incremental session. Hence, the model has strong classification and adaptability performance. In a word, our method can boost the model's stability for base classes and plasticity for incremental classes using the PITS and stochastic classifier, and thus achieves higher accuracy scores.

### *E. Statistical Significance Test*

From the results in last subsection, it is known that our method exceeds previous methods in accuracy. In this subsection, we discuss whether the accuracy advantage of our method is statistically significant. We do not know the distributions of samples in the three datasets. In addition, we do not assume that the samples in the datasets follows a specific distribution. A non-parametric method for statistical significance test is used, since it does not require the distribution of samples to be known. Specifically, the Friedman test [58] is used as the null-hypothesis test and then the Nemenyi test [58] is used as the post-hoc test for conducting the statistical significance test. The combination of the Friedman and Nemenyi tests is widely used in cases where the distribution of samples is unknown.

All methods are independently evaluated on different data subsets. The methods with the highest and the second highest AA scores on a data subset are ranked as 1 and 2, respectively, and so forth. Let $r_t^s$ represent the ranking obtained by the $s$th method on the $t$th data subset. The average ranking $R_s$ which is achieved by the $s$th method on $T$ data subsets is defined by

$$R_s = \frac{1}{T}\sum_{t=1}^{T} r_t^s \ . \tag{13}$$

where $1 \le s \le S$, $1 \le t \le T$; $S$ and $T$ denote total numbers of methods and data subsets, respectively. The definition of the null-hypothesis is that all methods have the same $R_s$ value, namely all methods have no statistically significant difference. If the result obtained from the Friedman test rejects the above null-hypothesis, then we use the Neminyi test to determine the statistically significant differences between all methods. If the difference in average rankings between two methods is larger than a critical difference (CD), then there is a statistically significant difference in performance between these two methods [58]. The CD is defined by

$$CD = q_\gamma \sqrt{\frac{S(S+1)}{6T}} \ . \tag{14}$$

where $q_\gamma$ is a critical coefficient which is determined based on the Studentized range statistic multiplied by $1/\sqrt{2}$ [58].

Our method and twelve previous methods are independently assessed using each of 30 data subsets. That is, $S$ = 13 and $T$ = 30. Each data subset is generated by randomly selecting samples from the FSC-89, NSynth-100, or LS-100. Each dataset is used to construct 10 data subsets. All methods are ranked according to their AA scores obtained on each data subset. As a result, each method has 30 rankings. The results of statistical significance test for all methods are shown in Fig. 5.

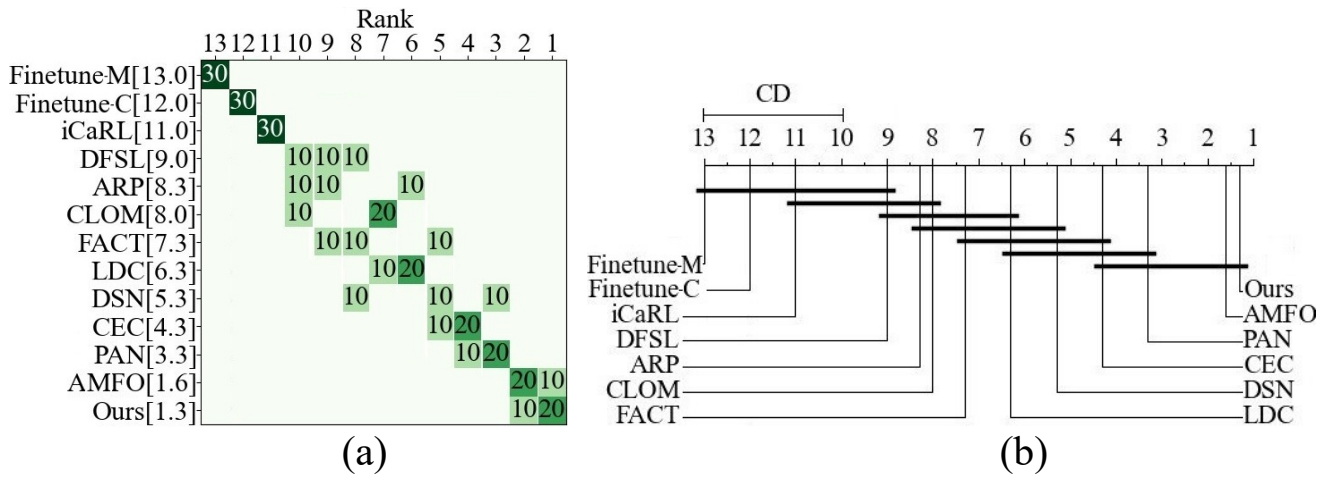

Fig. 5. Results of statistical significance test for all methods. (a) The number of times each method obtains each ranking. (b) The statistical significance differences for all methods.

The number of times each method achieves each ranking is shown in Fig. 5 (a). We sort all methods based on their average rankings (decimals in the square brackets on the left side of Fig. 5 (a)). The smaller the average ranking, the better the performance of the method. For example, our method is the best one among all methods since its average ranking of 1.3 is the smallest. The three best methods are ours, AMFO, and PAN, while the three worst methods are the Finetune-M, Finetune-C and iCaRL. The statistical significance differences for all methods are shown in Fig. 5 (b), where the confidence level $\gamma$ equals 0.05 (a typical setting). The horizontal axis denotes average rankings (from 1 to 13) for all methods. The performance difference between methods that are crossed by a black-thick crossbar is not statistically significant; otherwise, their performance difference is statistically significant. Our method has statistically significant advantage in accuracy over the methods of DSN, LDC, FACT, CLOM, ARP, DFSL, iCaRL, Finetune-C and Finetune-M, but does not statistically significantly exceed the methods of CEC, PAN and AMFO in accuracy.

### *F. Comparison in Complexity*

The model adopted in each method consists of an embedding learner and a classifier. The embedding learners of the models of various methods are structurally similar, but the classifiers are structurally different. The complexity difference between different methods is basically from their classifiers. The MACs, TT and IT are adopted to measure the computation complexity, while the NP is used to measure the memory overhead. Table IX presents the values of MACs, TT and IT of various methods on one sample from the LS-100, NSynth-100, and FSC-89. The values of NP of various methods on the three datasets are also given in Table IX. As the number of classes increases, the number of samples increases. Moreover, the training and inference time of various methods is directly proportional to the number of samples.

TABLE IX
VALUES OF MACs, TT/IT AND NP OF VARIOUS METHODS ON THREE DATASETS

| Method | MACs (million) | | | TT/IT (ms) | | | NP (million) | | |
|---|---|---|---|---|---|---|---|---|---|
| | LS | NS | FS | LS | NS | FS | LS | NS | FS |
| F-C | 310.00 | 601.52 | 278.70 | 28.84/1.38 | 33.02/1.58 | 38.25/1.83 | **11.18** | **11.18** | **11.18** |
| F-M | 310.00 | 601.52 | 278.70 | 28.84/1.38 | 33.02/1.58 | 38.25/1.83 | **11.18** | **11.18** | **11.18** |
| iCaRL | 310.02 | 601.55 | 278.73 | 28.63/1.37 | 30.93/1.48 | 37.62/1.80 | **11.18** | **11.18** | **11.18** |
| DFSL | 309.98 | 601.49 | 278.64 | 28.00/1.34 | 30.31/1.45 | 36.78/1.76 | 11.49 | 11.49 | 11.48 |
| CEC | 4560 | 9070 | 2310 | 108.89/5.21 | 118.71/5.68 | 142.54/6.82 | 12.28 | 12.28 | 12.27 |
| FACT | **299.13** | **579.86** | **267.64** | 26.96/1.29 | 29.26/1.40 | 36.37/1.74 | 11.22 | 11.22 | 11.22 |
| CLOM | 299.17 | 579.89 | 267.69 | 50.58/2.42 | 52.25/2.50 | 54.13/2.59 | 11.38 | 11.38 | 11.35 |
| LDC | 727.5 | 1008.21 | 696.07 | 95.51/4.57 | 100.32/4.80 | 128.74/6.16 | 12.79 | 12.79 | 12.79 |
| DSN | 301.48 | 582.47 | 269.53 | **22.15/1.06** | **23.41/1.12** | **34.28/1.64** | 13.58 | 13.84 | 13.05 |
| ARP | 1210 | 2320 | 1640 | 75.45/3.61 | 82.14/3.93 | 82.76/3.96 | 12.29 | 12.29 | 12.29 |
| PAN | 2620 | 2620 | 2330 | 76.97/4.72 | 78.73/4.83 | 105.78/6.49 | 13.33 | 13.33 | 13.33 |
| AMFO | 303.96 | 585.24 | 271.48 | 33.24/1.20 | 37.40/1.35 | 46.26/1.67 | 13.68 | 13.94 | 13.14 |
| Ours | 299.53 | 580.71 | 268.1 | 24.45/1.17 | 25.29/1.21 | 34.69/1.66 | 11.27 | 11.27 | 11.27 |

TT: Training Time; IT: Inference Time. NS: NSynth-100; FS: FSC-89; LS: LS-100. F-C: Finetune-C; F-M: Finetune-M.

From the MACs values of various methods, we can obtain the following three observations. First, the MACs values of our method are 299.53 million, 580.71 million and 268.1 million on the LS-100, NSynth-100 and FSC-89, respectively, and are smaller than their counterparts of previous methods except the methods of FACT and CLOM. Second, there are nine methods with similar and lower MAC values, including our method, AMFO, DSN, CLOM, FACT, DFSL, iCaRL, Finetune-M and Finetune-C. The main reason is that the classifiers used in these nine methods are structurally simple (such as fully-connected layers, prototypes) and do not have modules with high computational complexity. Third, the MACs values of the other four methods (PAN, ARP, LDC and CEC) are much larger than those of the above nine methods. This result is attributed to the fact that these four methods contain modules with high computational complexity. For example, the classifiers used in the methods of PAN, ARP and CEC contain attentive modules that are computationally complex. In the LDC method, the calibration of the covariance matrix which is used to update the classifier is computationally complex. From the TT and IT values of different methods, we can obtain observations and conclusions similar to the MACs values.

From the NP values of various methods, we can draw the following three conclusions. First, the NP value of our method is 11.27 million on all three datasets, and is smaller than its counterparts of previous methods except the methods of Finetune-M, Finetune-C, iCaRL and FACT. On the contrary, the AMFO method has the largest NP values on two datasets. The reasons for the above results are described as follows. First, the four methods with smaller NP values (including our method) use classifiers with simple structure and their structures do not expand with the increase of incremental classes. Second, the classifier of the AMFO method has three branches, whereas the classifiers adopted in other methods generally only consist of one branch.

Second, the NP values of eight methods are identical on three datasets, while the NP values of other five methods are not the same. The methods with the same NP values include Finetune-M, Finetune-C, iCaRL, FACT, LDC, ARP, PAN and ours, while the methods with different NP values include DFSL, CEC, CLOM, DSN, and AMFO. The main reason is that as the number of classes increases, the number of nodes in the last layer of the classifiers of the eight methods don't increase whereas the above node number of the other five methods will increase. The fewer classes, the smaller the NP values of these five methods; Otherwise, the bigger the NP values.

Third, the difference between the maximum and the minimum MACs values of different methods is 8802.36 (9070 - 267.64) million which is much larger than the difference of 2.76 (13.94 - 11.18) million between the maximum and the minimum NP values. The possible reason is that the embedding learners or classifiers with similar structures have comparable NP values but have prominently different MACs values. In addition, the embedding learners used in various methods are structurally similar, whereas the structures of classifiers are different.

In summary, sorting the complexity of the thirteen methods from low to high, our method ranks third, second and fifth in terms of MACs, TT/IT and NP, respectively. Moreover, the difference between the MACs value of our method and the smallest MACs value does not exceed 0.85 (580.71 - 579.86) million, while the difference between the NP value of our method and the lowest NP value is within 0.09 (11.27 - 11.18) million. In addition, the differences in TT and IT between our method and the lowest complexity method (namely the DSN) are below 2.3 (24.45 - 22.15) ms and 0.11 (1.17 - 1.06) ms, respectively.

In addition to comparing the complexity of various methods, we discuss the relationship between the NP values and AA scores of our method for showing that the setting of number of parameters for our method is reasonable. Since the embedding learner is the dominant part of the model in terms of NP, we only adjust the NP of the embedding learner here. Specifically, we simply increase or decrease the number of fully-connected layers of the embedding learner for adjusting the model's NP. As depicted in Fig. 6 (a), (b) and (c), our method obtains the highest AA scores when its NP value equals 11.27 million on all three datasets. When the NP values decrease or increase from 11.27 million, the AA scores of our method decline. The main reasons for the above results are explained below. When the NP value decreases from 11.27 million, the model size declines and the model will have difficulty in distinguishing various classes, especially base classes with abundant training samples. Therefore, the discriminative ability of the model becomes weaker. When the NP value increases from 11.27 million, the model size will rise and the model is inclined to overfit to various classes, especially incremental classes with few training samples. Hence, the generalization ability of the model will weaken. In conclusion, when the NP value of our method is 11.27 million on all three datasets, it achieves the highest accuracy scores and thus the parameter settings of our method are reasonable.

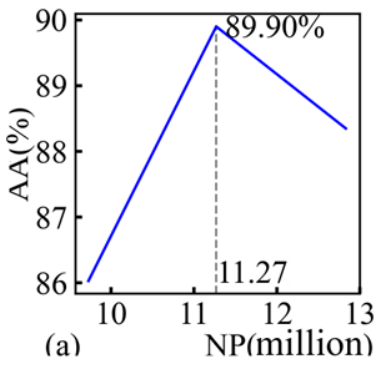

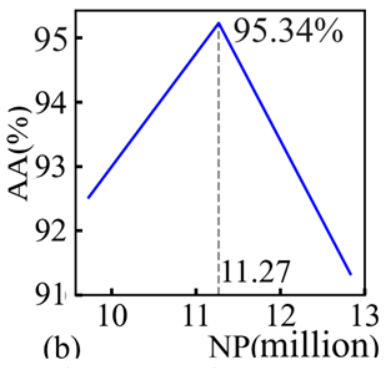

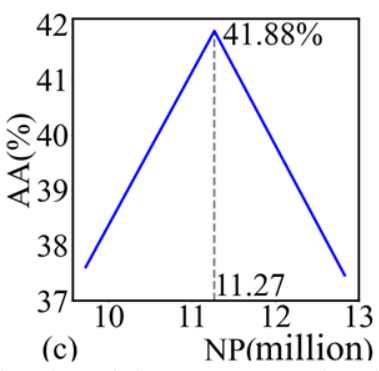

Fig. 6. Influence of the NP values on the AA scores obtained by our method on (a) LS-100, (b) NSynth-100 and (c) FSC-89.

### G. Comparison in Generalizability

We select training and testing samples from the same dataset in the experiments of all previous subsections. To compare the generalizability of various methods across different datasets, training and testing samples are chosen from two different datasets in the experiments of this subsection. The numbers of both way and shot are set to 5 (i.e., 5-way 5-shot) for few-shot learning. Table X presents the AA scores obtained by all methods on the combinations of two different datasets.

TABLE X
AA SCORES OBTAINED BY DIFFERENT METHODS ON THE COMBINATIONS OF TWO DIFFERENT DATASETS (IN %)

| Methods | FS→NS | FS→LS | NS→FS | NS→LS | LS→FS | LS→NS |
|---|---|---|---|---|---|---|
| Finetune-C [51] | 25.16 | 25.63 | 60.36 | 60.77 | 29.61 | 30.19 |
| Finetune-M [51] | 23.32 | 23.78 | 57.64 | 58.54 | 26.32 | 27.21 |
| iCaRL [52] | 26.53 | 25.83 | 73.52 | 72.63 | 42.97 | 41.86 |
| DFSL [29] | 32.08 | 33.17 | 81.12 | 78.05 | 73.26 | 74.04 |
| CEC [53] | 36.52 | 35.68 | 81.23 | 76.57 | 73.42 | 73.56 |
| FACT [54] | 39.45 | 32.21 | 82.53 | 78.14 | 77.88 | 74.91 |
| CLOM [55] | 38.71 | 35.68 | 75.84 | 76.48 | 73.60 | 76.18 |
| LDC [56] | 40.23 | 34.55 | 84.13 | 77.47 | 76.75 | 70.12 |
| DSN [57] | 34.63 | 31.74 | 83.09 | 75.00 | 66.45 | 67.80 |
| ARP [32] | 34.32 | 37.66 | 84.04 | 76.54 | 76.23 | 74.32 |
| PAN [30] | 40.31 | 39.37 | 83.83 | 77.50 | 78.71 | 77.31 |
| AMFO [33] | 41.22 | **45.79** | 84.21 | 78.25 | 78.27 | **77.59** |
| Ours | **42.41** | 41.96 | **84.37** | **78.29** | **78.74** | 76.42 |

NS: NSynth-100; FS: FSC-89; LS: LS-100.

As shown in the first row of Table X, the abbreviations on the left and right of each arrow denote training and testing datasets, respectively. For example, LS→FS denotes that the training dataset and testing dataset are LS-100 and FSC-89, respectively. Three datasets are used as training or testing data. Hence, there are six different combinations of datasets. The bottom row of Table X lists the AA scores obtained by our method on the six combinations of datasets. Among these six AA scores obtained by our method, four of them are higher than their counterparts obtained by other methods. Hence, in most cases, our method exceeds other methods in accuracy.

For the combinations of FS→FS, NS→NS and LS→LS, the training and testing samples are from the same dataset of FSC-89, NSynth-100 and LS-100, respectively. As shown in Tables VI, VII and VIII, the AA scores obtained by our method are 41.88%, 95.34% and 89.90% on the combinations of FS→FS, NS→NS and LS→LS, respectively.

The AA score of 41.88% on the combination of FS→FS is lower than 42.41% and 41.96% on the combinations of FS→NS and FS→LS, respectively. In addition, The AA score of 95.34% on the combination of NS→NS is higher than 84.37% on the NS→FS and 78.29% on the NS→LS. Similarly, the AA score of 89.90% on the combination of LS→LS is higher than 78.74% on the LS→FS and 76.42% on the LS→NS. It can be observed from the above AA scores that our method achieves higher AA scores when we select training and testing samples from the same datasets of both NSynth-100 and LS-100. In contrast, our method obtains lower AA scores when training and testing samples are chosen from the same dataset of FSC-89. The probable causes for the above results are explained as follows. The sample sources of the FSC-89 are diverse, and thus the distribution range of embedding space of the FSC-89 is very wide. In converse, the sample sources of the NSynth-100 and LS-100 are single, and thus the distribution ranges of their embedding space are very narrow. The embedding space of the FSC-89 probably overlaps greatly with the embedding space of the other two datasets, and may even cover the latter. When training samples are from the FSC-89, the trained model will have strong generalization ability on other datasets. Hence, our method obtains higher AA scores on the other two datasets.

In short, when the training and testing samples are from two different datasets, our method exceeds previous methods in most combinations of two datasets. It has good generalization ability across datasets, instead of overfitting to a single dataset.

### H. Extended Experiment

To show the performance of our method from more aspects, we conduct five extended experiments in this sub-section. In the first extended experiment, we discuss the impacts of various combinations of $N$ (number of ways) and $K$ (number of shots) on the AA scores obtained by our method. Without losing generality, only the LS-100 is used as the experiment dataset. Both $N$ and $K$ are set to 1, 5, 10 or 20. The accuracy scores in the last session are listed in Fig. 7 (a). Different settings of $N$-way and $K$-shot have impacts on the accuracy scores. The impacts have the following two characters.

First, under the same $N$, when $K$ increases, the accuracy scores steadily raise. Main reason is explained below. With the increase of $K$, more support samples can be used to update the stochastic classifier in each incremental session. Hence, the model can represent the incremental classes better and our method can obtain higher accuracy scores. When $K$ increases from 1 to 5, the minimum accuracy improvement is 6.91% (85.70% - 78.79%). The above improvement is larger than the maximum accuracy improvement, 1.96% (87.69% - 85.73%), when $K$ increases from 5 to 20. That is, when the number of support samples increases from 1 to 5, our method achieves a significant improvement in accuracy. However, when the number of support samples continues to increase from 5 to 20, the accuracy improvement is minor. The main reasons are explained below. One support sample cannot represent the characters of its class well, while five support samples have a strong ability to effectively represent the characters of their class. When the number of support samples increases from 5 to 20, there may be redundancy in the characters represented by each sample, and the accuracy improvement becomes minor.

Second, under the same $K$, the accuracy scores reach the maxima when $N$ is 5 and the accuracy scores decrease when $N$ deviates from 5, except in the case of 5-way 1-shot. The reasons are explained below. If the number of classes in each incremental session decreases, the number of incremental sessions increases. The probability of the model forgetting the knowledge of previous classes increases, which leads to a decrease in accuracy scores on previous classes. In contrast, if the number of classes in each incremental session increases, the number of incremental session decreases. It is hard for the model to accurately identify the classes in each incremental session. Therefore, confusions among various classes intensify and the accuracy scores decrease in each incremental session.

In the second extended experiment, we analyze the variance ranges of the stochastic classifier for various classes. Without losing generality, the distribution of twenty means and variances are illustrated in Fig. 7 (b) using the t-SNE [59]. The means and variances of the twenty classes are randomly selected from the LS-100. The means and variances of these classes don't overlap, and the variance of each class is closest to the mean of the class it belongs to. The variance of each class is closely centered around its corresponding mean.

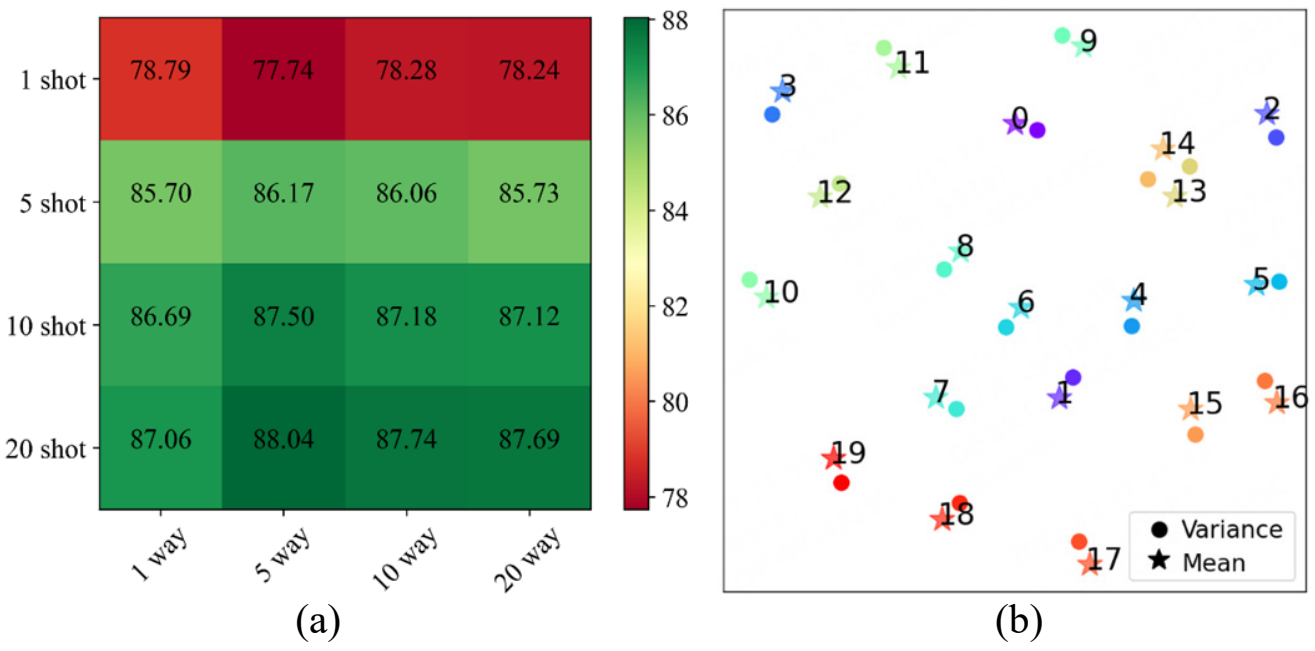

Fig. 7. (a) Accuracy scores obtained by our method in the last session under different settings of $N$-way $K$-shot (in %); (b) Visualization of twenty means and variances of the stochastic classifier.

In the third extended experiment, we discuss the impacts of $\alpha$ (a parameter of the Beta distribution) on the AA scores obtained by our method. $\alpha$ influences the probability of $\lambda$ (mixing factor) taking different values, and thus controls the mixed results of samples and labels of the pseudo-incremental classes. The AA scores achieved by our method with different $\alpha$ values on the FSC-89, NSynth-100 and LS-100 are shown in Fig. 8 (a), (b) and (c), respectively. When $\alpha$ equals 0.4, 3.0 and 4.0, our method obtains the highest AA scores on the FSC-89, NSynth-100 and LS-100, respectively. When $\alpha$ changes, the variation in AA scores on the three datasets is less than 0.50% (42.00% - 41.50%). Hence, our method is robust to the values of $\alpha$. In addition, there are no failure cases where the tuning of $\alpha$ fails in our experiments. Based on the results in Fig. 8, the setting of $\alpha$ values can range from 0 to 5.0 across datasets.

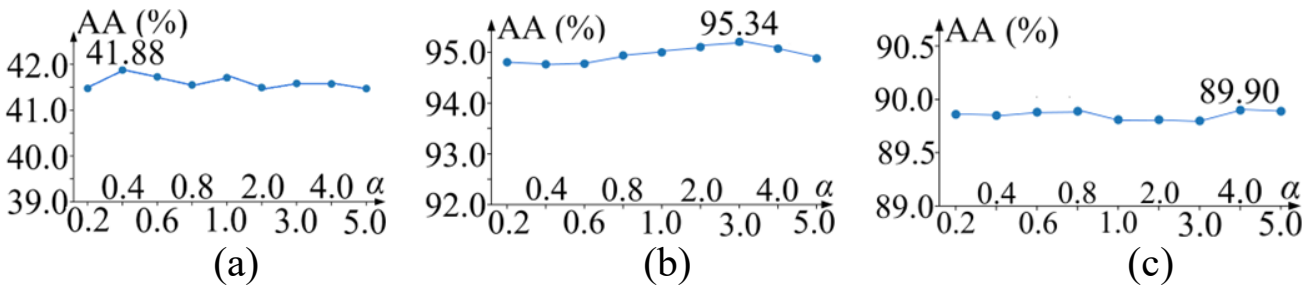

Fig. 8. The impact of $\alpha$ on the AA scores obtained by our method on the (a) FSC-89; (b) NSynth-100; and (c) LS-100. The values of the abscissa are not proportional.

In the fourth extended experiment, we discuss the impacts of $M'$ (number of pseudo-incremental sessions) on the AA scores obtained by our method. The AA scores achieved by our method with different $M'$ values on the FSC-89, NSynth-100 and LS-100 are shown in Fig. 9 (a), (b) and (c), respectively. When $M'$ is equal to 6, 9 and 8, our method obtains the highest AA scores on the FSC-89, NSynth-100 and LS-100, respectively. When $M'$ changes, the variation in AA scores on the three datasets is less than 0.60% (95.34% - 94.74%). Hence, our method is robust to the values of $M'$. In addition, there are no failure cases where the tuning of $M'$ fails in our experiments. Based on the results in Fig. 9, the setting of $M'$ values can range from 1 to 10 across datasets.

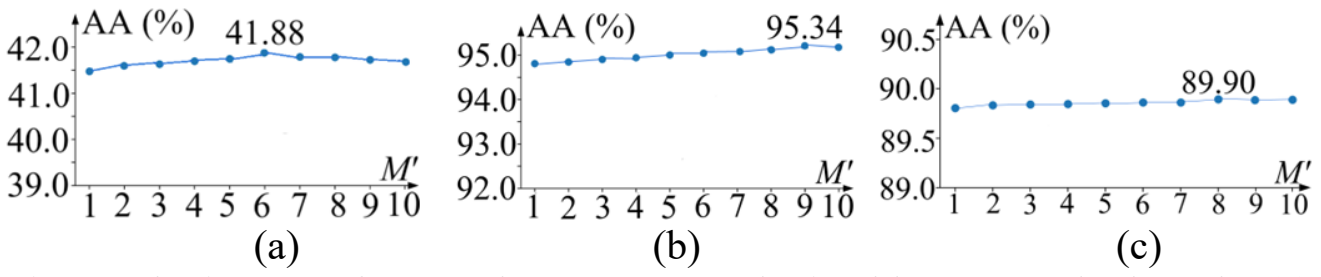

Fig. 9. The impact of $M'$ on the AA scores obtained by our method on the (a) FSC-89; (b) NSynth-100; and (c) LS-100.

In the fifth extended experiment, we discuss feature space adaptation through plotting the embeddings before and after performing the PITS. Without losing generality, we randomly select 10 base classes and 5 pseudo-incremental classes. Fig. 10 (a) shows the embeddings of the 10 base classes, while Fig. 10 (b) illustrates the embeddings of the 10 base classes and 5 pseudo-incremental classes. It can be observed from Fig. 10 that the embeddings of these randomly-selected classes are separated from each other in the feature space, rather than being mixed together. The result in Fig. 10 indicates that the PITS is able to generate embeddings of pseudo-incremental classes that are far apart from embeddings of base classes in the feature space.

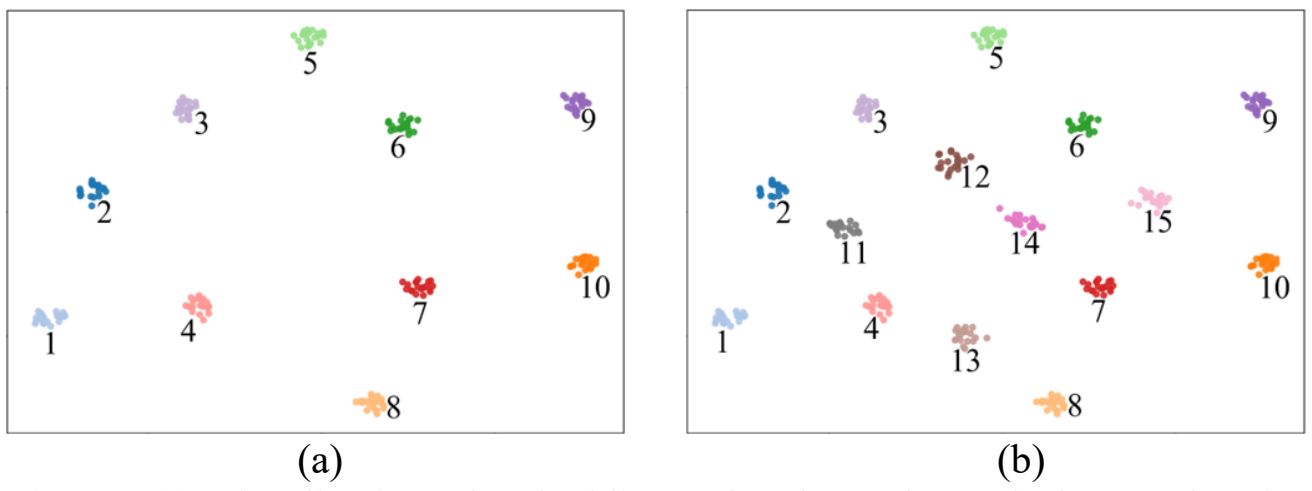

Fig. 10. (a) Visualization of embeddings of 10 base classes before performing the PITS; (b) Visualization of embeddings of both 10 base classes and 5 pseudo-incremental classes after performing the PITS.

## IV. Conclusions

In this paper, we proposed a FCAC method with a pseudo-incrementally trained embedding learner and a continually updated stochastic classifier. To enable the embedding learner to possess strong representation ability, we proposed the PITS to train the embedding learner with augmented samples and labels. To enable the stochastic classifier to own strong classification ability, each weight of the stochastic classifier is represented by a distribution which consists of a mean vector combined with a variance vector.

We can draw four conclusions based on the description of our method and results. First, our method almost achieved the highest AA scores and lowest PD values on all datasets among all methods. It exceeded most of prior methods in accuracy with statistical significance. Second, among all methods, our method ranks third, second and fifth in terms of MACs, TT/IT and NP, respectively. Third, our method basically exceeded all previous methods when the training and testing samples are from different datasets. It had strong generalization ability across datasets, instead of overfitting to a single dataset. Fourth, our method is robust to the values of main hyperparameters.

Our method achieved better results than prior methods. However, there are still some issues worth exploring in order to further improve the performance of our method or enhance the completeness of our work. First, the structure of the embedding learner needs to be improved. The embedding learner is structurally a convolutional neural network that can represent the short-term characters of each sample well, but lacks the ability to represent long-term characters. Hence, we will consider integrating self-attention or mutual-attention modules into the embedding learner since they can represent long-term characteristics well. Second, we will take effective measures (e.g., model distillation, parameter compression) to further reduce the complexity of our method for deploying it on edge devices, such as smart-speakers and audio monitors.

Therefore, we will implement our method on the edge devices for few-shot class-incremental recognition of speech keywords, anomalous sound of devices, animal calls, and so on. Third, we will consider using other distributions (such as Laplace distribution) to generate the weights of stochastic classifier and then compare the performance of various distributions. Hence, we will provide more ways of weights generation for stochastic classifiers.

## References

[1] Y. Li, Q. Wang, X. Li, X. Zhang, Y. Zhang, A. Chen, Q. He, and Q. Huang, "Unsupervised detection of acoustic events using information bottleneck principle," *Digital Signal Processing*, vol.63, pp.123-134, 2017.

[2] Z. Lin, et al., "Domestic activities clustering from audio recordings using convolutional capsule autoencoder network," in *Proc. of IEEE ICASSP*, 2021, pp. 835-839.

[3] Y. Li, X. Li, Y. Zhang, M. Liu, and W. Wang, "Anomalous sound detection using deep audio representation and a BLSTM network for audio surveillance of roads," *IEEE Access*, vol. 6, pp. 58043-58055, 2018.

[4] Y. Li, M. Liu, K. Drossos, and T. Virtanen, "Sound event detection via dilated convolutional recurrent neural networks," in *Proc. of IEEE ICASSP*, 2020, pp. 286-290.

[5] J. Tan, and Y. Li, "Low-complexity acoustic scene classification using blueprint separable convolution and knowledge distillation," in *Technical Report of DCASE2023 Challenge,* 2023, pp. 1-4.

[6] Y. Li, W. Cao, W. Xie, Q. Huang, W. Pang, and Q. He, "Low-complexity acoustic scene classification using data augmentation and lightweight ResNet," in *Proc. of IEEE ICSP*, 2022, pp. 41-45.

[7] W. Xie, Q. He, H. Yan, and Y. Li, "Acoustic scene classification using deep CNNs with time-frequency representations," in *Proc. of IEEE ICCT*, 2021, vol. 4, pp. 1325-1329.

[8] Y. Li, J. Tan, G. Chen, J. Li, Y. Si, and Q. He, "Low-complexity acoustic scene classification using parallel attention-convolution network", in *Proc. of Interspeech*, 2024, pp. 1-5.

[9] W. Pang, Q. He, Y. Li, and N. Ahmed, "Detecting video anomalies by jointly utilizing appearance and skeleton information," *Expert Systems with Applications*, vol. 246, pp. 1-12, 2024.

[10] W. Pang, et al., "Audiovisual dependency attention for violence detection in videos," *IEEE TMM*, vol. 25, pp. 4922-4932, 2023.

[11] Y. Li, W. Wang, M. Liu, Z. Jiang, and Q. He, "Speaker clustering by co-optimizing deep representation learning and cluster estimation," *IEEE TMM*, vol. 23, pp. 3377-3387, 2021.

[12] Y. Chen, et al., "Interrelate training and clustering for online speaker diarization," *IEEE/ACM TASLP*, vol. 32, pp. 1352-1364, 2024.

[13] Y. Li, Z. Jiang, Q. Huang, W. Cao, and J. Li, "Lightweight speaker verification using transformation module with feature grouping and fusion," *IEEE/ACM TASLP*, vol. 32, pp. 794-806, 2024.

[14] Y. Li, H. Chen, W. Cao, Q. Huang, and Q. He, "Few-shot speaker identification using lightweight prototypical network with feature grouping and interaction," *IEEE TMM*, vol. 25, pp. 9241-9253, 2023.

[15] M. Scarpiniti, et al., "Deep recurrent neural networks for audio classification in construction sites," in *Proc. of EUSIPCO*, 2021, pp. 810-814.

[16] L. Zhang, Z. Shi, and J. Han, "Pyramidal temporal pooling with discriminative mapping for audio classification," *IEEE/ACM TASLP*, vol. 28, pp. 770-784, 2020.

[17] A. Quelennec, M. Olvera, G. Peeters, and S. Essid, "On the choice of the optimal temporal support for audio classification with pre-trained embeddings," in *Proc. of IEEE ICASSP*, 2024, pp. 976-980.

[18] Y. Hou, et al., "Multi-level graph learning for audio event classification and human-perceived annoyance rating prediction," in *Proc. of IEEE ICASSP*, 2024, pp. 716-720.

[19] T. Alex, S. Ahmed, A. Mustafa, M. Awais, and P. J. Jackson, "Max-AST: combining convolution, local and global self-attentions for audio event classification," in *Proc. of IEEE ICASSP*, 2024, pp. 1061-1065.

[20] S. Shishkin, D. Hollosi, S. Goetze, and S. Doclo, "Active learning for sound event classification using bayesian neural networks with gaussian variational posterior," in *Proc. of IEEE ICASSP*, 2024, pp. 896-900.

[21] Y. Sun, et al., "Automated data augmentation for audio classification," *IEEE/ACM TASLP*, vol. 32, pp. 2716-2728, 2024.

[22] Z. Bai, C. Pan, G. Chen, J. Chenk, and J. Benesty, "A weighted binary cross-entropy for sound event representation learning and few-shot classification," in *Proc. of APSIPA ASC*, 2023, pp. 1069-1074.

[23] X. Zhang, et al., "Frame-level embedding learning for few-shot bioacoustic event detection," in *Proc. of IEEE ICME*, 2023, pp. 750-755.

[24] I. Moummad, N. Farrugia, and R. Serizel, "Regularized contrastive pre-training for few-shot bioacoustic sound detection," in *Proc. of IEEE ICASSP*, 2024, pp. 1436-1440.

[25] J. Liang, H. Phan, and E. Benetos, "Learning from taxonomy: multi-label few-shot classification for everyday sound recognition," in *Proc. of IEEE ICASSP*, 2024, pp. 771-775.

[26] Z. Wang, et al., "Continual learning of new sound classes using generative replay," in *Proc. of WASPAA*, 2019, pp. 308-312.

[27] M. Mulimani, and A. Mesaros, "Class-incremental learning for multi-label audio classification," in *Proc. of IEEE ICASSP*, 2024, pp. 916-920.

[28] M.A. Hussain, et al., "An Efficient Incremental Learning Algorithm for Sound Classification," *IEEE MultiMedia*, vol. 30, no. 1, pp. 84-90, 2023.

[29] Y. Wang, et al., "Few-shot continual learning for audio classification," in *Proc. of IEEE ICASSP*, 2021, pp. 321-325.

[30] Y. Li, et al., "Few-shot class-incremental audio classification using dynamically expanded classifier with self-attention modified prototypes," *IEEE TMM*, vol. 26, pp. 1346-1360, 2024.

[31] W. Xie, Y. Li, Q. He, and W. Cao, "Few-shot class-incremental audio classification via discriminative prototype learning," *Expert Systems with Applications*, vol. 225, 120044, pp. 1-13, 2023.

[32] W. Xie, et al., "Few-shot class-incremental audio classification using adaptively-refined prototypes," in *Proc. of Interspeech*, 2023, pp.301-305.

[33] Y. Li, J. Li, Y. Si, J. Tan, and Q. He, "Few-shot class-incremental audio classification with adaptive mitigation of forgetting and overfitting," *IEEE/ACM TASLP*, vol. 32, pp. 2297-2311, 2024.

[34] T. Azevedo, R.d. Jong, M. Mattina, and P. Maji, "Stochastic-YOLO: Efficient probabilistic object detection under dataset shifts," in *Proc. of Workshop on Machine Learning for Autonomous Driving at the NeurIPS*, 2020, pp. 1-9.

[35] C. Cheng, L. Zhang, H. Li, L. Dai, and W. Cui, "A deep stochastic adaptive Fourier decomposition network for hyperspectral image classification," *IEEE TIP*, vol. 33, pp. 1080-1094, 2024.

[36] H. Zhang, M. Cisse, Y.N. Dauphin, D. Lopez-Paz, "Mixup: beyond empirical risk minimization," in *Proc. of ICLR*, 2018, pp. 1-13.

[37] Z. Lu, et al., "Stochastic classifiers for unsupervised domain adaptation," in *Proc. of IEEE/CVF CVPR*, 2020, pp. 9108-9117.

[38] Y. Li, et al., "Few-shot class-incremental audio classification using stochastic classifier," in *Proc. of Interspeech*, 2023, pp. 4174-4178.

[39] Y. Li, M. Liu, W. Wang, Y. Zhang, and Q. He, "Acoustic scene clustering using joint optimization of deep embedding learning and clustering iteration," *IEEE TMM*, vol. 22, no. 6, pp. 1385-1394, 2020.

[40] Y. Li, et al., "Domestic activity clustering from audio via depthwise separable convolutional autoencoder network," in *Proc. IEEE 24th Int. Workshop Multimedia Signal Process.*, 2022, pp. 1-6.

[41] W. Xie, et al., "Deep mutual attention network for acoustic scene classification," *Digital Signal Processing*, vol. 123, 2022, Art. no. 103450.

[42] Y. Li, et al., "Speaker verification using attentive multi-scale convolutional recurrent network," *Appl. Soft Comput.*, vol. 126, 2022, Art. no. 109291.

[43] K. He, X. Zhang, S. Ren, and J. Sun, "Deep residual learning for image recognition," in *Proc. of IEEE/CVF CVPR*, 2016, pp. 770-778.

[44] Z. Duan, et al., "QARV: Quantization-aware ResNet VAE for lossy image compression," *IEEE TPAMI*, vol. 46, no. 1, pp. 436-450, 2024.

[45] Q. Wang, et al., "A four-stage data augmentation approach to ResNet-Conformer based acoustic modeling for sound event localization and detection," *IEEE/ACM TASLP*, vol. 31, pp. 1251-1264, 2023.

[46] M. Magris, and A. Iosifdis, "Bayesian learning for neural networks: an algorithmic survey," *Artificial Intelligence Review*, vol. 56, no. 10, pp. 11773-11823, 2023.

[47] L.V. Jospin, et al., "Hands-on Bayesian neural networks—A tutorial for deep learning users," *IEEE Computational* Intelligence Magazine, vol. 17, no. 2, pp. 29-48, 2022.

[48] Y. Wang, N. J. Bryan, J. Salamon, M. Cartwright, and J. P. Bello, "Who calls the shots? Rethinking few-shot learning for audio," in *Proc. of IEEE WASPAA*, 2021, pp. 36-40.

[49] J. Engel, C. Resnick, A. Roberts, S. Dieleman, M. Norouzi, D. Eck, and K. Simonyan, "Neural audio synthesis of musical notes with WaveNet autoencoders," in *Proc. of ICML*, 2017, pp. 1068-1077.

[50] V. Panayotov, G. Chen, D. Povey, and S. Khudanpur, "Librispeech: An ASR corpus based on public domain audio books," in *Proc. of IEEE ICASSP*, 2015, pp. 5206-5210.

[51] P. Dhar, R.V. Singh, K.-C. Peng, Z. Wu, and R. Chellappa, "Learning without memorizing," in *Proc. IEEE/CVF CVPR*, 2019, pp. 5138-5146.

[52] S.A. Rebuffi, et al., "iCaRL: Incremental classifier and representation learning," in *Proc. of IEEE CVPR*, 2017, pp. 5533-5542.
[53] C. Zhang, et al., "Few-shot incremental learning with continually evolved classifiers," in *Proc. of IEEE/CVF CVPR*, 2021, pp. 12450-12459.
[54] D.W. Zhou, et al., "Forward compatible few-shot class-incremental learning," in *Proc. of IEEE/CVF CVPR*, 2022, pp. 9036-9046.
[55] Y. Zou, et al., "Margin-based few-shot class-incremental learning with class-level overfitting mitigation," in *Proc. of NeurIPS*, 2022, pp. 1-13.
[56] B. Liu, et al., "Learnable distribution calibration for few-shot class-incremental learning," *IEEE TPAMI*, vol. 45, no. 10, pp. 12699-12706, 2023.
[57] B. Yang, et al., "Dynamic support network for few-shot class incremental learning," *IEEE TPAMI*, vol. 45, no. 3, pp. 2945-2951, 2023.
[58] J. Demšar, "Statistical comparisons of classifiers over multiple data sets," *Journal of Machine Learning Research*, vol.7, pp. 1-30, 2006.
[59] L.V.D. Maaten, and G. Hinton, "Visualizing data using T-SNE," *J. Mach. Learn. Res.*, vol. 9, no. 96, pp. 2579-2605, 2008.